\documentclass[prodmode]{acmsmall} 
\pdfoutput=1

\usepackage[ruled]{algorithm2e}
\usepackage{comment}
\usepackage[T1]{fontenc}
\usepackage[latin1]{inputenc}

\usepackage{makecell}
\usepackage{multirow}
\usepackage{graphics}

\usepackage{balance}

\usepackage{epsfig} 
\usepackage{mathptmx}
\usepackage{color}
\usepackage{float}
\usepackage{cite}
\usepackage{subfigure}

\usepackage{amsmath}
\usepackage{amsfonts}
\usepackage{amssymb}
\usepackage{bm}

\usepackage[
  pass,
]{geometry}

\SetAlFnt{\small}
\SetAlCapFnt{\small}
\SetAlCapNameFnt{\small}
\SetAlCapHSkip{0pt}
\IncMargin{-\parindent}

\newcommand{\squeezeup}{\vspace{-2.5mm}}

\begin{document}

\markboth{Tajalizadehkhoob et al.}{What We Are Measuring When We Are Measuring Abuse}

\title{Rotten Apples or Bad Harvest? What We Are Measuring When We Are Measuring Abuse}
\author{SAMANEH TAJALIZADEHKHOOB
\affil{Delft University of Technology}
RAINER {B\"{O}HME}
\affil{Innsbruck University}
CARLOS GA\~N\'AN
\affil{Delft University of Technology}
MACIEJ KORCZY\'NSKI
\affil{Delft University of Technology}
MICHEL VAN EETEN
\affil{Delft University of Technology}
}

\begin{abstract}
Internet security and technology policy research regularly uses technical indicators of abuse in order to identify culprits and to tailor mitigation strategies. As a major obstacle, readily available data are often misaligned with actual information needs. They are subject to measurement errors relating to observation, aggregation, attribution, and various sources of heterogeneity. More precise indicators such as size estimates are costly to measure at Internet scale.
We address these issues for the case of hosting providers with a statistical model of the abuse data generation process, using phishing sites in hosting networks as a case study. We decompose error sources and then estimate key parameters of the model, controlling for heterogeneity in size and business model. We find that 84\,\% of the variation in abuse counts across 45,358 hosting providers can be explained with structural factors alone. Informed by the fitted model, we systematically select and enrich a subset of 105 homogeneous ``statistical twins'' with additional explanatory variables, unreasonable to collect for \emph{all} hosting providers. We find that abuse is positively associated with the popularity of websites hosted and with the prevalence of popular content management systems. Moreover, hosting providers who charge higher prices (after controlling for level differences between countries) witness less abuse. These factors together explain a further 77\,\% of the remaining variation, calling into question premature inferences from raw abuse indicators on security efforts of actors, and suggesting the adoption of similar analysis frameworks in all domains where network measurement aims at informing technology policy. 
 
 \end{abstract}

\maketitle




\section{Introduction}
\label{sec:intro}

Abuse data is an important foundation for security and policy research.
It associates technical identifiers -- typically IP addresses, domain names or URLs -- with malicious activities, such as spam, infected machines, command-and-control servers, and phishing sites.

Scientific studies and industry reports draw on abuse data to make inferences about the security efforts of the parties in charge of the networks or services where the abuse is located. Concentrations of abuse are seen as evidence of poor security practices or even criminal behavior, explicitly or implicitly characterizing certain providers as ``rotten apples'', or at least as actors who can and should do more remediation \cite
{trendmicro2015,hao2013understanding,czyz2014taming,stone2009fire,shue2012abnormally,arman}.

Industry representatives often counter these kind of inferences when they make media headlines. 
For example, a 2013 McAfee report ranked the Netherlands as number three worldwide in hosting botnet command--and--control (C\&C) servers~\cite{mcafee13}. 
A leading news organization concluded: ``Netherlands Paradise for Cybercriminals'', prompting a debate that reached the national parliament~\cite{nos13}. 
The Dutch Hosting Provider Association responded that it ``disagrees vehemently'' with this conclusion~\cite{dhpa}. 
It argued that the high ranking was an artifact of the large hosting infrastructure in the country, not of any negligence or malice on the part of providers. 

The hosting provider association raised a valid point. 
We know that concentrations of abuse are, to a large extent, a function of the size of the network and the service portfolio of the provider, rather than of the provider's security effort \cite{clayton2015}.
More customers, for example, mean higher numbers of compromised accounts. 

At the heart of such disagreements is the issue of discretion: the degree in which provider efforts drive their abuse levels. Or conversely: the degree in which abuse levels indicate the security efforts of a provider.
The core question of this paper is to what extent abuse levels are determined by the security efforts of individual providers versus being a function of structural features of the industry.
More colloquially: are abuse levels driven by rotten apples or an overall bad harvest?

A major obstacle for making valid inferences from abuse measurements is the fact that the technical indicators are often misaligned with actual information needs. 
They are subject to measurement errors relating to observation, aggregation, attribution, and various sources of heterogeneity such an unreliable size estimates. Whilst the fact that size -- as one of the many providers' structural properties -- matters sounds rather obvious, the exiting body of work have not yet controlled for this effect in a systematic and statistically sound manner.
Most of the related studies who study the concentrations of abuse in a network, often correlate or normalize abuse with a size estimate as a proxy for the number of customers, whilst such methods often fall short if one does not 
control for other structural or security-effort related provider properties~\cite{stone2009fire,shue2012abnormally}.

Therefore, to advance our ability for making more reliable inferences from abuse data, in this paper, we develop an analytical approach and propose a statistical model of the abuse data generation process to understand 
to what extent are abuse levels determined by the security efforts of individual providers versus being a function of structural properties of the industry?
We use phishing abuse data as a case-study to demonstrate our approach.
In short, we make the following contributions:

\begin{itemize}
\item We present a scalable approach to study abuse across the population of hosting providers, as opposed to technical entities such as Autonomous Systems, using passive DNS and \texttt{WHOIS} data;

\item We replace existing regression models to explain abuse with a more sophisticated modeling approach that carefully decomposes different sources of variance at work in the generation of abuse data;

\item Our approach is able to distinguish security efforts from structural properties of providers when explaining concentrations of abuse in networks. In a case study on phishing data, we show that over 84\% of the variance in abuse data can be explained by the structural properties of providers, such as their size and business model. This implies that security efforts only have a minor impact on abuse counts;


\item We develop an approach using 'statistical twins' to measure the impact of factors that are difficult to measure at scale. Using this approach, we present the first empirical evidence of the impact of pricing, time-in-business and the amount of Wordpress sites on phishing abuse. Together with other factors related to provider business models, we are able explain a further 77\,\% of the remaining variation in abuse, while controlling for country-level effects using fixed-effect model;

\item We demonstrate how our approach generates comparative abuse metrics by controlling for the structural differences among providers. Such relative metrics are more suited to evaluate countermeasures than absolute counts and concentrations of abuse.

\end{itemize}

Measurements of the absolute amount of abuse across providers are clearly relevant, assuming they are attributed correctly to the appropriate agent. They fall short, however, when establishing agency, developing interventions and holding individual providers accountable. 
Our approach is an alternative that allows actors to take into account a wider range of factors and support more informative inferences from abuse data for security and policy issues.

In the following Section \ref{sec:approach}, we outline the analytical approach that sets up the rest of the paper.
Section \ref{sec:datasets} describes data sources and collection methodology.
Section \ref{sec:regsize} describes our modeling approach and interprets the results.
In Section \ref{sec:pairs}, we add additional provider structural properties for sample of enriched data points.
In Section \ref{sec:measure_error} we evaluate the impact of measurements errors around abuse data and size estimates. 
Section \ref{sec:related_work} covers the related work. Finally, we discusses our main conclusions and implications in Section \ref
{sec:conclusions}.

\section{Related Work}
\label{sec:related_work}

There is an enormous amount of work on methods for detecting abused resources. The blacklists based on these detections have also been extensively studied \cite{pitsillidis2012taster,kuhrer2014paint}. 
Observational data on abuse incidents is the starting point for our study. We do not engage with the detection methods themselves and therefore will not survey them here.

Closer to our research is the rich body of work that identifies and explains patterns in abuse data. The explanations are situated at different levels of analysis: (i) individual resource (host, IP address, domain); (ii) network, or other aggregates of individual resources; (iii) actor, meaning the economic entity providing the resources or otherwise responsible for them; and (iv) country. We briefly survey relevant work at each level.

\textbf{Individual resources.}
A variety of studies have been very effective at explaining or predicting the occurrence of abuse, such as compromised websites, from the properties of individual resources, such as webserver software versions \cite{vasek2015hacking,soska2014automatically}. 
The factors identified in these studies impact the distribution of abuse across providers, e.g.,
if customers of a provider use popular software with known vulnerabilities, then the abuse rate will go up. 
Others used DNS-specific characteristics to predict whether domain names are malicious \cite{bilge2011exposure,hao2011monitoring}.
More tailored towards phishing, some methods propose to detect phishing websites using URL and content-specific features \cite{rosiello2007layout,whittaker2010large}.

\textbf{Networks.} Another strand of work looks at how abuse is distributed across aggregate units of resources, such as address blocks, Autonomous Systems (ASes) \cite{levchenko2011click,arman} or  top-level domains (TLDs) \cite{apwg2015}.
These studies often identify concentrations of abuse in certain networks \cite{ramachandran2006understanding,collins2007using,levchenko2011click} and then identify network features that correlate with abuse rates,
such as poor network hygiene
\cite{zhang2014mismanagement,stone2009fire} or rapidly changing network connectivity \cite{shue2012abnormally,wagner2013asmatra,Konte2015aswatch}. 
These studies aim to detect malicious or poorly managed networks, rather than disentangling the factors explaining the causal relationship of why abuse is concentrated there or how it is distributed across all networks. Furthermore, to be useful for interventions and policy, the aggregated resources need to be attributed to the relevant actor.
For drawing inferences about providers, an explicit attribution methodology needs to be in place. 
This takes us to the third level of analysis.

\textbf{Actors.} Actors are the economic entities that operate the resources or are otherwise responsible for them. 
Work at this level has to bridge the gap between the technical identifiers in abuse data, like 
Autonomous Systems (ASes) and the organizations responsible for those resources. 
This is hardly straightforward since many 
Internet Service Providers (ISPs), for example, operate multiple ASes \cite{asghari2015economics} and many hosting providers, on the other hand, operate with on average seven other providers in the same AS \cite{nomshosting2016}.

Our work is situated at this level.
We are not aware of any work that explains abuse patterns across hosting providers. 
Somewhat related is a study of security practices in a small sample of providers \cite{canali2013role}.
More mature work is available for the domain name space. Studies have identified which registrar or registries are associated with malicious domains and how they could intervene \cite{liu2011effects,hao2013understanding,levchenko2011click}.

\textbf{Countries.} The highest level of analysis is countries. 
Work in this area studies the relationship between country-level factors, such as GDP, 
rule of law and ICT development, and the distribution of abuse, most notably infected hosts \cite{asghari2015post,garg2013macroeconomic,wang2009cyber,kleiner2013linking}.

Our study extends the prior work in a variety of ways.
Despite the existing work, we are not trying to find technical features to correlate with abuse, but we are trying to understand 
to what extent are abuse levels determined by the security efforts of individual providers versus being a function of structural properties of the industry?
For that, first we adopt a better attribution method for identifying the relevant actors, by moving from ASes to hosting providers, to whom the IP space is allocated.  
We restrict our analysis of abuse to phishing instances, as a case study, and we fit a advanced statistical model to grasp the extent to which we can explain the reason behind heterogeneity observed in abuse counts of hosting providers. Our model includes a set of important explanatory  factors, some of which such as size of IP space has been partially  explored before, and others such as portion of shared hosting business, hosting price, time in business and the amount of Wordpress sites which are studied for the first time in our paper. In addition, we advanced our analysis by controlling for country-level effects in our models.

\section{Analytical Approach}
\label{sec:approach}

\begin{figure}
	\centering
\includegraphics[scale=0.5]{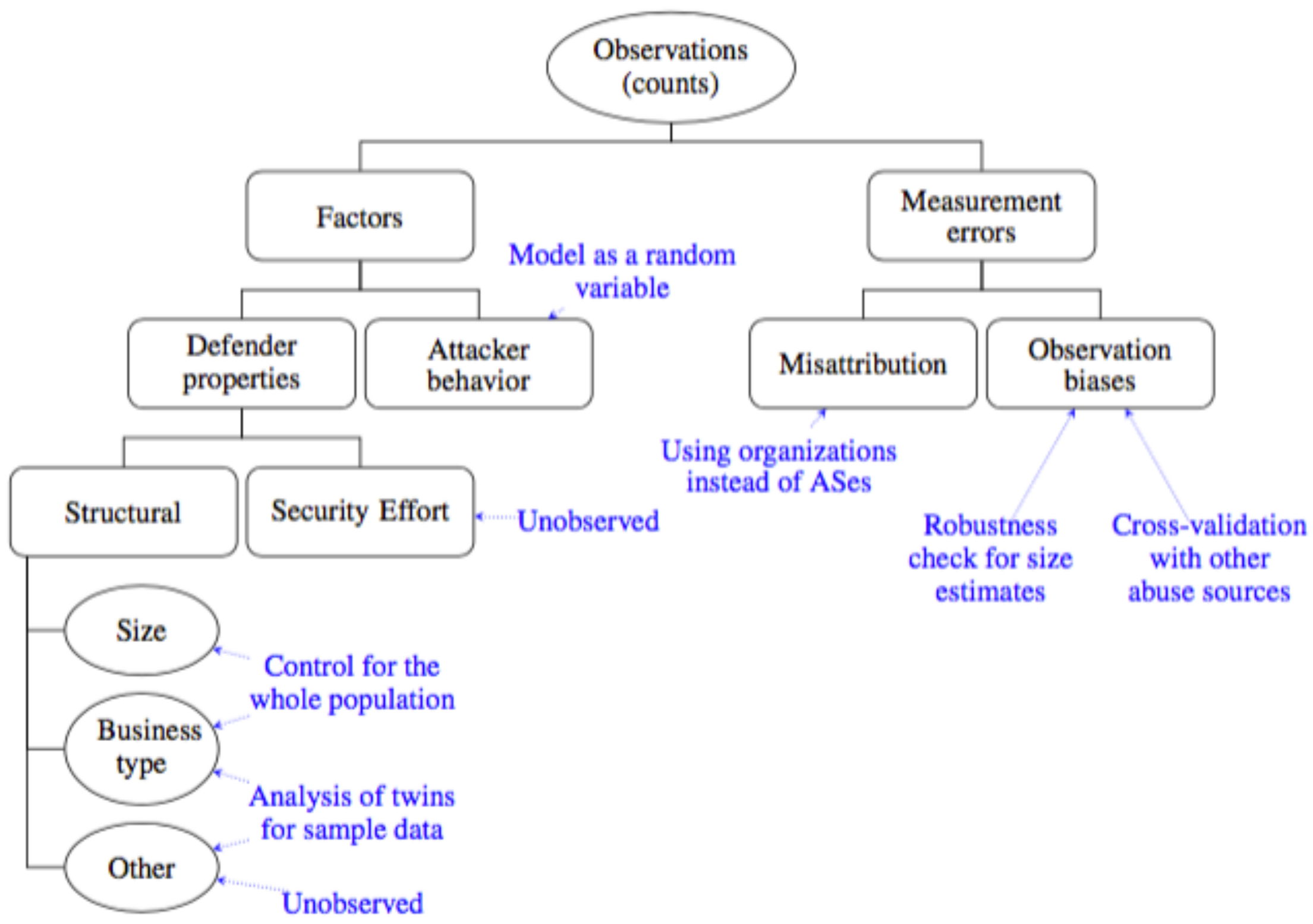}
\label{fig:anal_model}
\caption{Overview of our analytical approach}
\end{figure}

To assess the driving forces behind abuse levels of hosting providers, we need to disentangle a variety of factors and errors at work in abuse counts.
Figure \ref{fig:anal_model} summarizes these sources of variance and provides an overview of our approach.
We assume that abuse observations are driven by two phenomena: explanatory factors (left) and measurement errors (right).

We divide the explanatory factors into defender properties and attacker behavior.
Defender properties are then branched to two main groups: structural and security effort related properties.
Relevant structural properties of defender include the size of the provider, as recognized in earlier work \cite{clayton2015,arman,stone2009fire,Konte2015aswatch}.
Simply put, the more domains or IP addresses a provider hosts, the higher the odds of incurring an abuse incident, as there is a larger number of potential victims in the network. 

In addition to size, the business type is relevant. 
Providers offer different portfolios of hosting services, e.g., unmanaged, dedicated, shared, Virtual Private Server, etc. 
These services differ in the extent to which the provider is responsible or in a position to provision security \cite{M3AAWG}.
Business type also includes the pricing strategy. 
Other characteristics include the legal framework under which providers operate and the overall maturity of ICT infrastructure development in a country.
Last, but not least, the security efforts of the providers will influence the abuse rate.

Next to explanatory factors, we distinguish measurement errors that cause variance in the abuse counts: biases in observations (for factors and abuse data) and problems in attributing abuse incidents to the responsible entity.

To find out to what extent defender properties explain the variance in the abuse counts of hosting providers, we develop a statistical modeling approach and implement it for a one type of abuse data as a case-study: phishing domains.
The blue text in Figure~\ref{fig:anal_model} indicates how we deal with the various sources of variance decomposed by the analytical model. 
We first apply Poisson Generalized Linear Models (GLMs) to the number of phishing domains per provider, and estimate the coefficients for four structural defender properties related to size and business type 
(detailed in Section 4).
These variables can be economically collected for all hosting providers.

Some additional structural defender properties are more cumbersome to collect, since they require manual work or are costly to others when measured at scale. A statistician's response would be to estimate from a random sample. The size of a random sample depends on the target level of confidence and on the effect size (akin the signal-to-noise ratio). In the case of hosting providers, the heterogeneity in the population may hide subtle effects of security effort, which would require uneconomically large samples to control for. 

A more efficient approach is to modify the sampling strategy and select subsets of cases which appear homogeneous according to the observable defender properties. Specifically, we select a set of ``statistical twins'' -- subsets of size two -- covering the domain of the known population. We collect additional variables for each twin. The subsequent analysis looks for factors explaining differences \emph{within} twins, disregarding differences \emph{between} twin. Technically, this can be achieved by adding one fixed effect per subset to the GLM specification. This method allows us to control for large parts of the heterogeneity and at the same time account for linear bias introduced by the systematic sample selection. 

Another way of looking at this approach is that we select the a priori most informative cases from the population for further analysis. It rests on the implied assumption that cases which have a twin in the population do not systematically differ from those which do not. We assume that this is plausible for the population of global hosting providers.

Of course, there are always further factors that remain unobserved.
The same holds for the security efforts of providers. We have no direct observations of these efforts, which means that they are part of the unexplained variance in our model. 
All unobserved factors, such as missing defender properties, attacker behavior, together with all measurement errors, are conflated in the unexplained variance of the model. We establish the boundaries of the impact that all these unobserved factors have. In other words, we show that even though there is a lot we cannot observe, the few things we can observe explain a big portion of the variance in the abuse counts.


We follow the former interpretation and account for it as measurement error that we can attenuate, if not avoid completely. Specifically, we reduce attribution errors in identifying hosting providers by replacing routing data (BGP), the typical basis for attribution, with \texttt{WHOIS} to map domain names and IP addresses in passive DNS data to their corresponding hosting providers. 
Avoiding observational biases is outside the scope of this paper, but we limit their effect on our core results by cross-validation against a different set of phishing data. Finally, we test the robustness of the size estimates in the model against errors due to possible model mismatch by simulating the impact of distorted defender properties against a simulated phishing count variable drawn from an ideal Poisson distribution model.

\section{Data}
\label{sec:datasets}
Our study deploys a variety of data sources, which are summarized in Table~\ref{tab:sum_vars}.

\subsection{Hosting provider identification}
We want to accurately identify which IP ranges belong to actual hosting providers.
Most of the existing work uses BGP data to map IP addresses of abuse incidents to ASes, equating the latter with hosting providers.
However, the entity that is routing an IP address is not always the same as the entity hosting an IP address.  
\texttt{WHOIS} data provides more direct visibility into who is responsible for an IP address. In should be noted that \texttt{WHOIS} data comes with its well-known limitations such as different data formats, they are less detrimental on analysis of the hosting market than starting with BGP \cite{elliott2008and}.
Cai et al. \cite{cai2010towards} describe a general method for mapping IP addresses to their organizations using \texttt{WHOIS}. 
We refined this method for the hosting market.

In a nutshell, we only map organizations which host domains seen in DNSDB, a passive DNS database that draws upon hundreds of sensors worldwide and which is generously provided to us by Farsight Security \cite{farsight,dnsdb}.
The data contains second-level domain names and the IP addresses that they resolved to. 
To our knowledge, DNSDB has the best coverage of the domain name space available to researchers.
We take the IP addresses and domains seen in DNSDB in 2015 and map them to their corresponding organizations using \texttt{WHOIS} data and the MaxMind API \cite{maxmind,liu2015learning}.
To compile a final set of organizations who offer hosting services, we adopt some of the keywords and categories suggested in \cite{nomshosting2016,dimitropoulos2006revealing} and then exclude educational and government-related organizations from our set.
The final set consists of 45,358 hosting providers.

\subsection{Abuse data}
In order to answer the main question of the paper, we model the count of abuse in the networks of hosting providers, using phishing data as a case-study.
The main reason behind this choice is, since phishing sites are known to be mostly compromised accounts \cite{apwg2015}, 
bypassing security is very much required in the bulk of cases.
To that end, we analyze phishing domains collected from two sources: the Anti-Phishing Working Group (APWG)~\cite{apwg} and Cyscon GmbH~\cite{cyscon}. 

\textbf{APWG data} contains domains of phishing domains from all over the world,
blacklisting times, and other meta-data.
We collect the IP addresses associated with second-level domains\footnote{Domains such as \texttt{example.co.uk} are considered to be second-level domains as well.} in the APWG feed by retrieving the corresponding passive DNS entry at the time when the domain is blacklisted.
The final set consists of 131,018 unique second-level domains and 95,294 unique IP addresses for the whole year 2015. 

\textbf{Cyscon phishing data} has the same attributes (URLs, IP addresses, blacklisting times, meta-data).
We collect 40,292 second-level domains and 23,021 unique IP addresses for June--December 2015.

We use the phishing second-level domains in APWG data as the default response variable in our regression models in Section \ref{sec:regsize} and \ref{sec:pairs}. In Section~\ref{sec:measure_error}, we use the Cyscon data to cross-validate the results.

\subsection{Provider properties}
To explain the differences in phishing incident counts between hosting providers, we collect a number of variables on provider (defender) properties, some for the entire population and some for the sample of ``statistical twins''.

\subsubsection{Variables collected for all providers}
\label{subsec:data_popoulation}

In addition to identifying providers, we can collect variables related to size and business model (see the leftmost factors in Fig.~\ref{fig:anal_model}) from passive DNS and \texttt{WHOIS} data.

\textbf{Number of assigned IP addresses.} 
Size of IP address block(s) assigned to a provider based on \texttt{WHOIS}.

\textbf{Number of IP addresses hosting domains.} 
Count of IP addresses seen to host domains in passive DNS.
The combination of these two variables provides information about the kind of business the hosting provider is running. 
For instance, providers who use a large part of their assigned IP space for hosting domains such as webhosting providers can have a different business model from providers who use their IP space for hosting other services such as data centers.

\textbf{Number of hosted domains.} 
Count of the second-level domains in the passive DNS data. In addition, note that since the first three variables have a skewed distribution, we log-transform them with base 10.

\textbf{Percentage of domains hosted on shared IPs.} We consider an IP address shared, if it hosts more than 10 domain names \cite{nomshosting2016}. 

This variable measures the ratio of domains that are hosted on shared IP addresses over the total size of the hosted domains, in percent.
This variable not only conveys information about the size of the shared hosting infrastructure of the provider, but also about the provider's business model: the degree to which it relies on low-cost shared hosting services. 

\squeezeup 
\begin{table}[!htbp] 
\centering 
\tbl{Descriptive statistics of variables used in the full model for all providers and providers in the sample of twins\label{tab:sum_vars} }{%
\resizebox{0.65\linewidth}{!}{
\begin{tabular}{lrrrrrr} 
\\[-1.8ex]\hline 
\hline \\[-1.8ex] 
& \textbf{min} & \textbf{mean} & \textbf{median}& \textbf{max} & \textbf{sd} \\ 
\hline \\[-1.8ex] 
 \textbf{for all data points} (n = $45,358$)   & &  &  &  &  \\ 
 \cline{1-1}
 \\[-1.8ex]
 \# assigned IPs [log10] & $0$ & $3.1$ & $3.2$ & $8.4$ & $1.2$ \\ 
\# IPs hosting domains [log10] & $0$ & $1.8$ & $1.7$ & $6.2$ & $0.8$ \\ 
\#  hosted domains [log10] & $0$ & $2.0$ & $1.8$ & $7.6$ & $0.9$ \\ 
\% domains hosted on shared IPs & $0$ & $51.0$ & $59.0$ & $100$ & $37.1$ \\[1ex]
\# phishing domains in APWG & $0$ & $2.8$ & $0$ & $11,455$ & $91.3$ \\ 
\# phishing domains in Cyscon  & $0$ & $0.9$ & $0$ & $5,515$ & $37.4$ \\ 
\\[-1ex]
\textbf{for statistical twins} (n = $210$) & &  &  &  &  \\ 
 \cline{1-1}
 \\[-1.8ex]
\# assigned IPs [log10] & $0.3$ & $4.0$ & $4.0$ & $7.5$ & $1.4$ \\ 
\# IPs hosting domains [log10] & $0.3$ & $3.0$ & $3.0$ & $5.6$ & $1.2$ \\ 
\#  hosted domains [log10] & $1.5$ & $3.9$ & $3.7$ & $7.6$ & $1.2$ \\ 
\% domains hosted on shared IPs & $0$ & $78.6$ & $87.9$ & $99.3$ & $22.3$ \\[1ex] 
\# phishing domains in APWG & $0$ & $159.2$ & $3$ & $9,805$ & $967.6$ \\ 
\# phishing domains in Cyscon & $0$ & $54.8$ & $1$ & $3,819$ & $375.0$ \\ 
\hline \\[-1.8ex] 
\end{tabular} }}
\squeezeup 
\end{table}

\subsubsection{``Statistical twins'' sampling method}
\label{subsec:sampling}

It is not possible or desirable to collect data at scale for all factors in Figure~\ref{fig:anal_model}. For example, pricing information must be collected manually. It involves search, interpretation, and human judgment. Applying a standardized procedure is too costly for the entire population of 45,358 hosting providers for some of these variables.
Even collecting some technical indicators, such as the number of Wordpress installation on \emph{all} websites of \emph{every} hosting provider, are inefficient to collect in bulk; perhaps even unethical, because it imposes a cost on the scanned networks.

For this reason, we employ a data-driven sampling strategy and select a small set of homogeneous ``statistical twins'' for which we can freely  collect as much information as possible. The steps are as follows:

\begin{itemize}
\item [(i)] We start with a set of randomly and uniformly selected \textit{seed data points} (105 hosting providers) for which we have collected pricing information.
Let $\mathcal{S}$ be the set of seed data points and $\mathcal{T}$ be the total set (or population) of providers. The random seed should ensure a good coverage of the population.
\item [(ii)] 
We calculate the distance between all the data points in $\mathcal{S}$ and data points in $\mathcal{T}$ using the Euclidean distance between all explanatory variables collected for the entire population. 
This results in a distance matrix of 105 rows and around 45\,k columns.
\item [(iii)] For each of the 105 providers in $\mathcal{S}$, we select the closest match; that is the provider in set $\mathcal{T}$ that has the minimum distance to the provider in set $\mathcal{S}$, in terms of variables in Table \ref{tab:sum_vars}.
\item [(iv)] This results in a set of \emph{rich data points} $\mathcal{R}$ consisting of 105 homogeneous statistical twins and 206 unique hosting providers in total, where a few providers became part of two twins.
We further use set $\mathcal{R}$ as our (notoriously biased) sample to study the effect of additional factors on abuse. We account for the bias in the analysis (Section \ref{sec:measure_error}).
\item [(v)] We collect additional variables for all elements in  $\mathcal{R}$. 
\end{itemize}

While the method is economical and increases the information gained per effort, it comes with some limitations. First, it is unlikely to spot outliers simply because there is nothing to pair up against an extreme hosting provider such as {Amazon}. Second, it assumes that data points in dense regions of the population do not systematically differ from those in sparser regions, where the probability of finding twins is lower. Third, it requires that the bias introduced by the selection strategy is approximately linear. Non-linear bias correction is possible in principle, but requires prior information on the functional form.

\begin{figure}[ht!]
\squeezeup 
\centering
\includegraphics[width=.35\columnwidth]{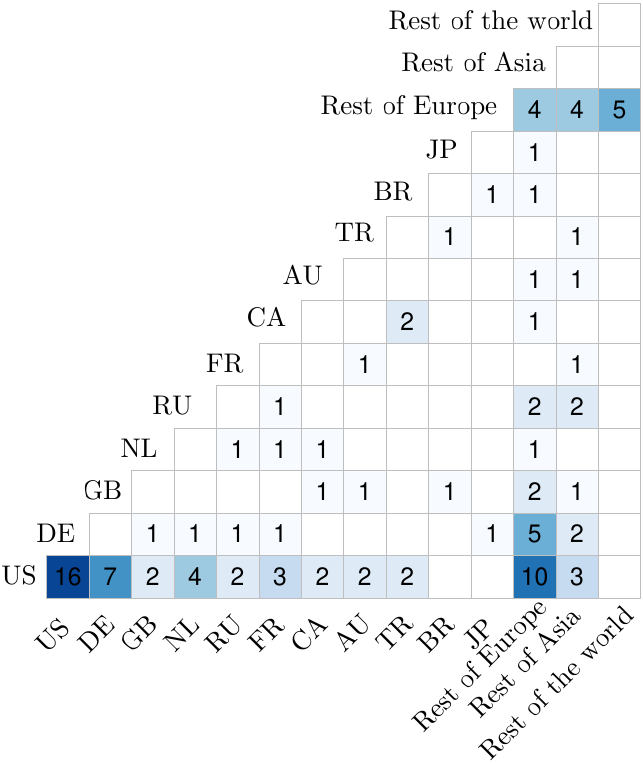}
\caption{Number of twins per country combination}
\label{fig:pairs_cc}
\squeezeup 
\end{figure}

\subsubsection{Variables collected for a sample of providers}
\label{subsec:data_sample}
The below variables are collected only for the providers in $\mathcal{R}$, as explained in Section \ref{subsec:sampling}. 

\textbf{Country.}
For the providers in $\mathcal{R}$, we use MaxMind API to identify where the majority of IP addresses are located. In 
Figure \ref{fig:pairs_cc}, we show the general coverage of twins in our data. For instance, there are 16 twins with both providers located in US while there is 1 twin with one provider located in Britain and the other in Japan. The figure indicates a good variation in the geo-location of twins in our sample data.

\textbf{ICT development index.} The ITU publishes coun\-try-level variables that proxy the development in information and communication technology (ICT) using several indicators \cite{Ictindex}. (This and the previous variable came into view for our enriched data points, but they could also be economically collected for the whole population.) 

\textbf{Popularity index.}
We use Alexa's one million top-ranked domains as a proxy for online popularity.
A provider is assumed to host more popular content when more domains are on the list. 
If attacker behavior is not completely random, one may expect that providers with more popular sites are targeted more in order to compromise domains and set up phishing sites. 

The popularity index per provider in calculated by aggregating the Alexa ranks of second-level domains in our sample $\mathcal{R}$.
We first reverse the Alexa ranks per domain, that is the most popular Alexa domain gets the rank 1,000,000. We then calculate a score per provider by summing up the base-10 logarithm of the reverse rank of all ranked websites. 
This score combines website (i.e., customer) popularity and the density of popular sites at a hosting provider and the method allows us to account for extreme popularity of large providers.

\textbf{Time in business}.
Time is business is a proxy for how experienced a provider is.
We collect this information by querying the \texttt{WHOIS} database for the registration date of the provider's own domain name.
Missing values were entered if we could not find the website or there was no public registration date in \texttt{WHOIS}.
We cross-checked the results with the Internet Archive for all data points \cite{webarchive}.
Almost all domains in our sample were captured by Web-archive a couple of months after the day they were registered.

\textbf{Pricing}.
Finding comparable pricing information for hosting providers is complicated. We visit the provider's website and collect prices for the least expensive hosting plan on offer, as an indicator to tell `bottom-end' from `top-end' apart.
We hypothesize that providers with cheaper hosting plans have fewer resources and more vulnerable customers, so higher phishing counts.
We converted all prices to US dollars by taking the 2015 average exchange rate of the local currency to USD.
Price information for providers is missing if (i) we could not find the provider's website; (ii) the prices are not available online; and (iii) 
we do not receive a reply to our inquiries through other channels.

\begin{figure*}[!htbp]
\squeezeup 
  \centering
  \subfigure[ICT development index]{\includegraphics[width = .32\textwidth]{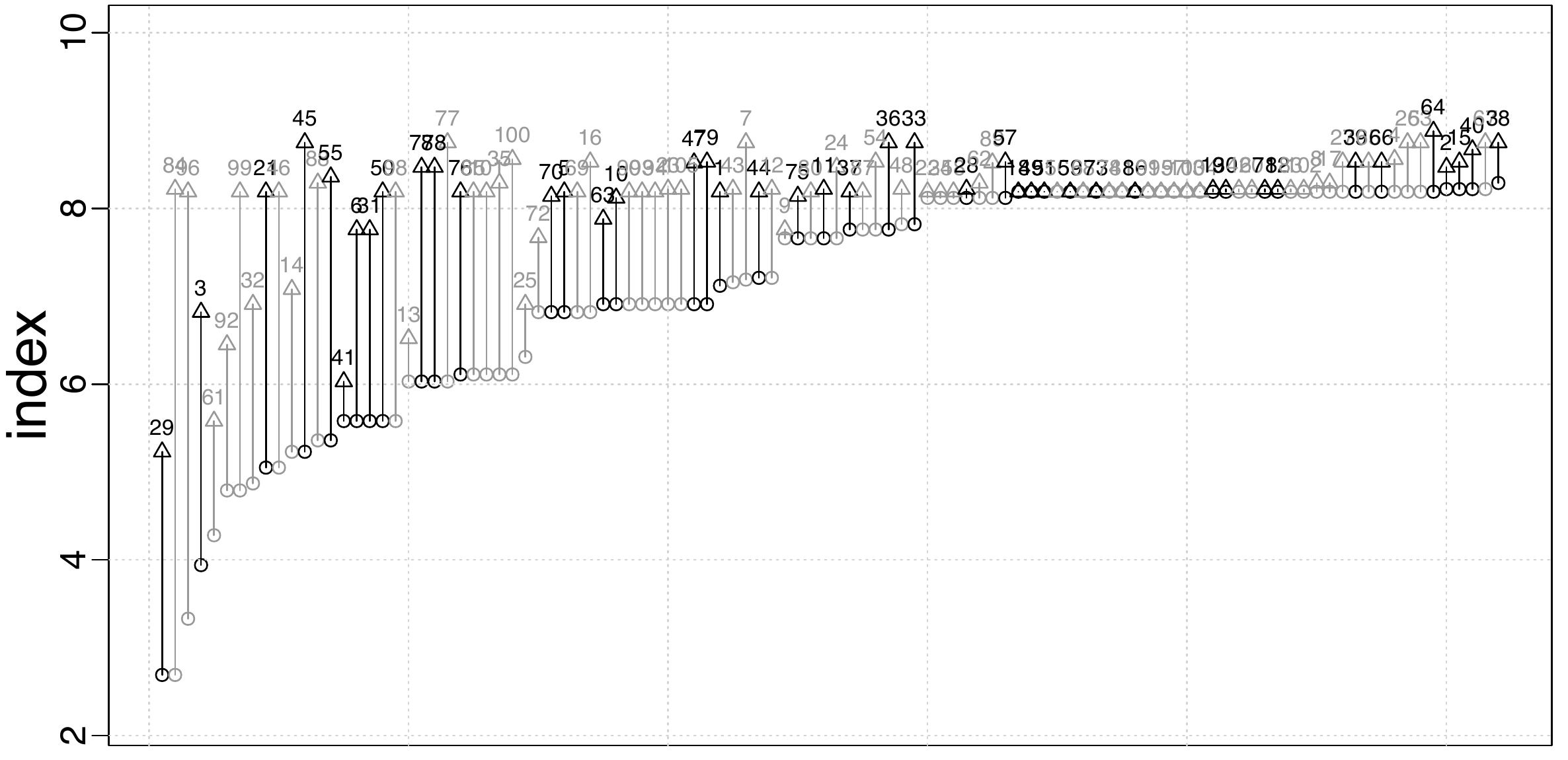}\label{fig:pairs-ict}}
  \subfigure[Popularity index]{\includegraphics[width = .32\textwidth]{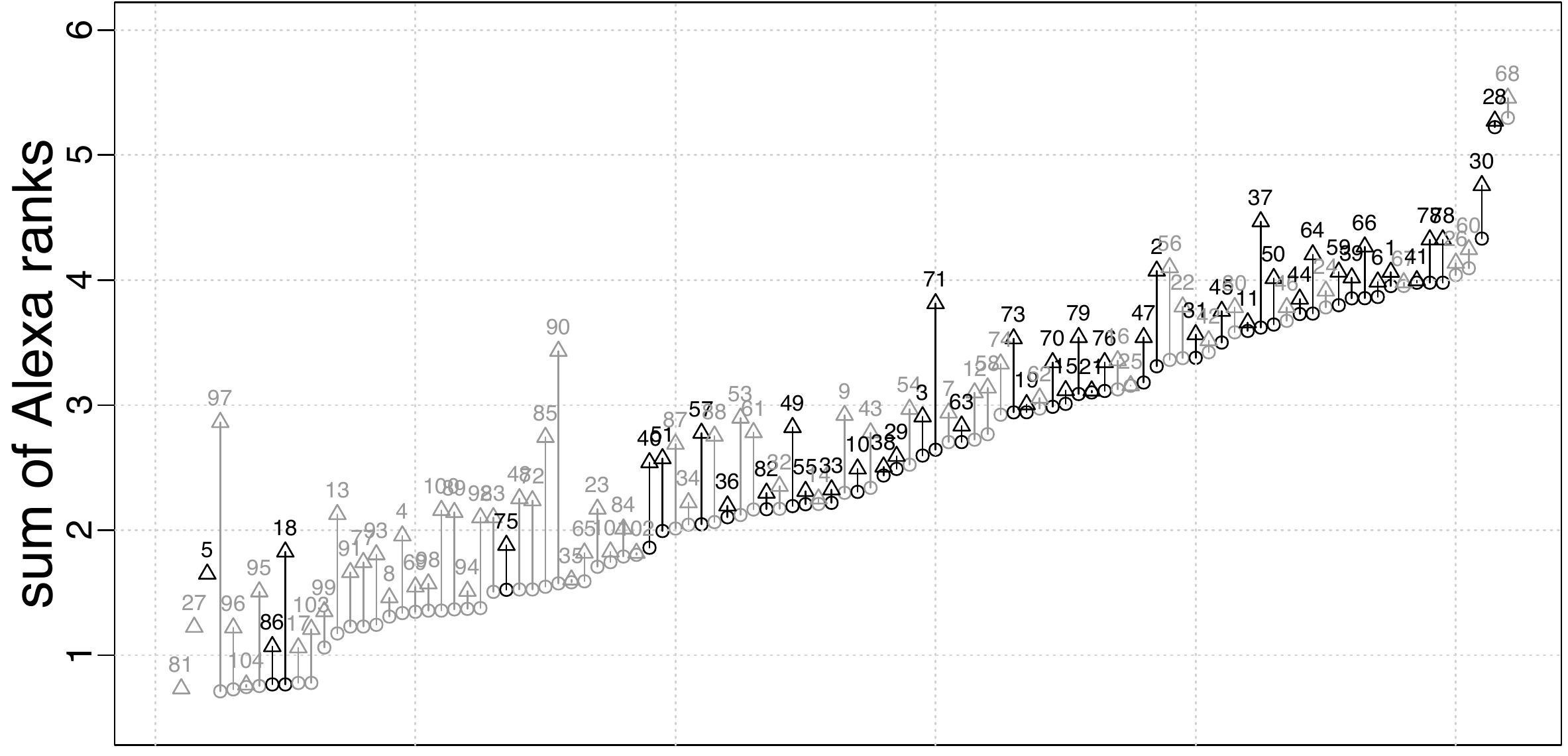}\label{fig:pairs-alexa}}
  \centering
  \subfigure[Time in business]{\includegraphics[width = .32\textwidth]{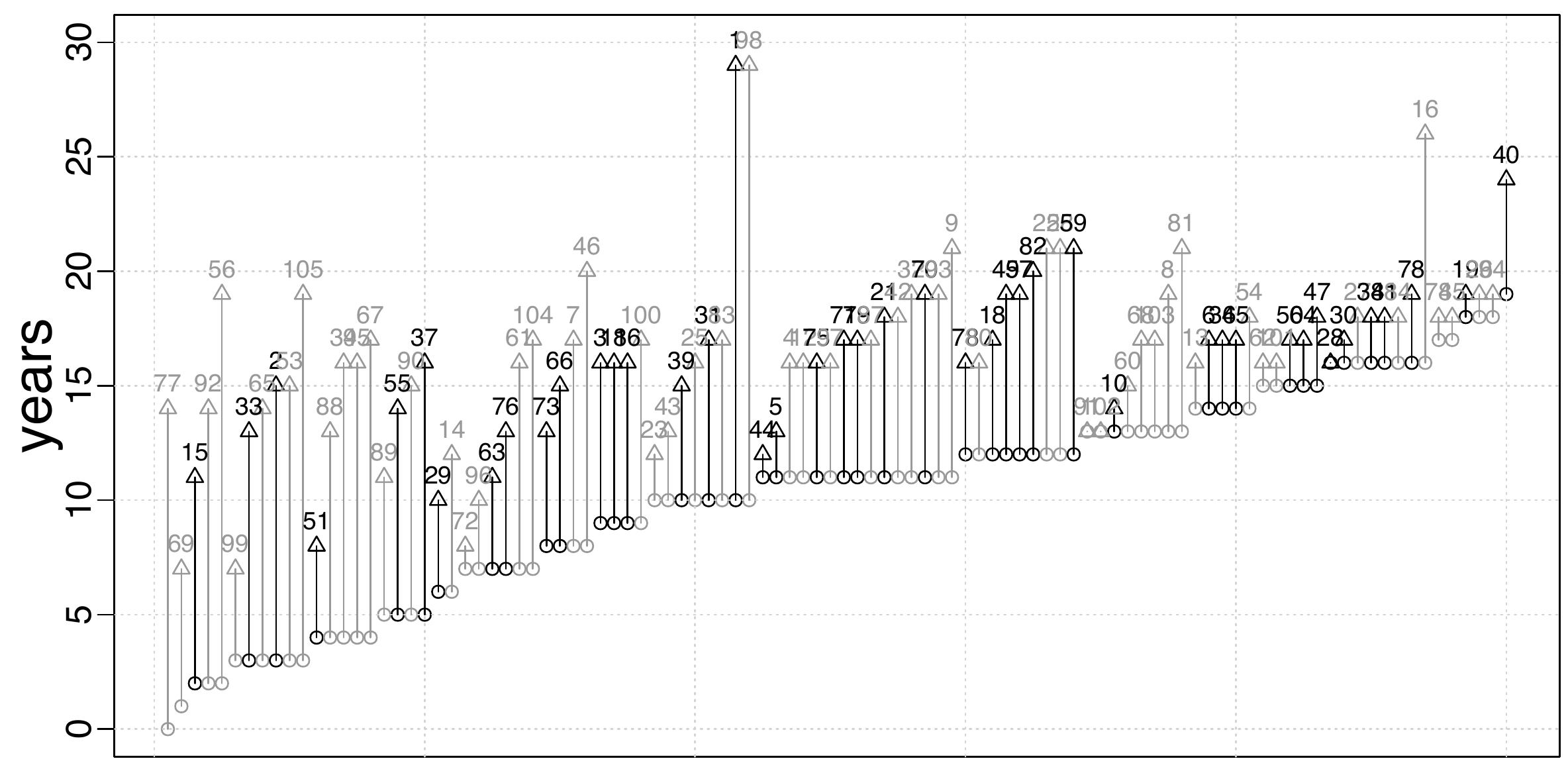}\label{fig:pairs-tib}}
\subfigure[Price per year]{\includegraphics[width = .32\textwidth]{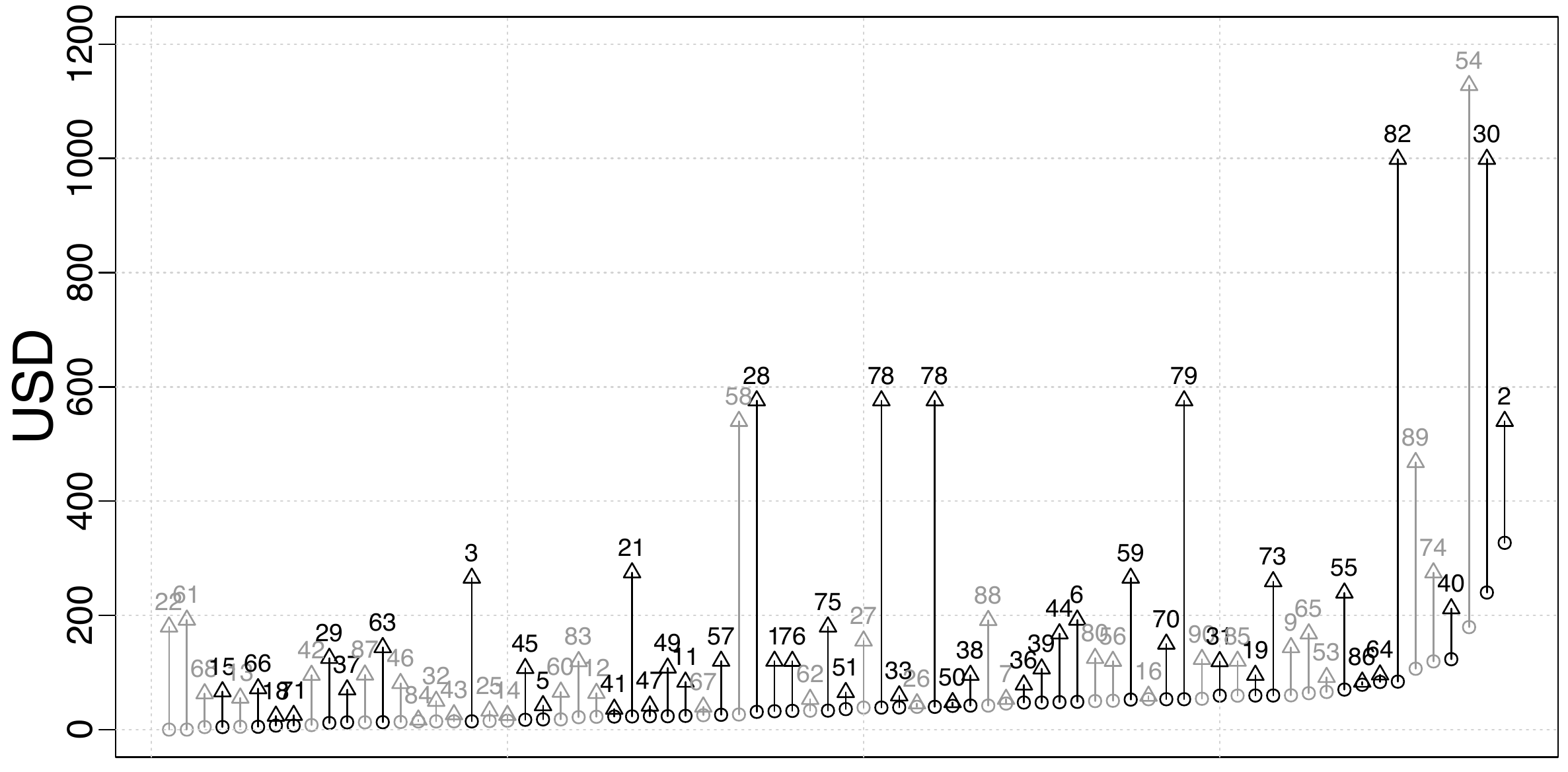}\label{fig:pairs-price}}
\subfigure[Wordpress use]{\includegraphics[width = .32\textwidth]{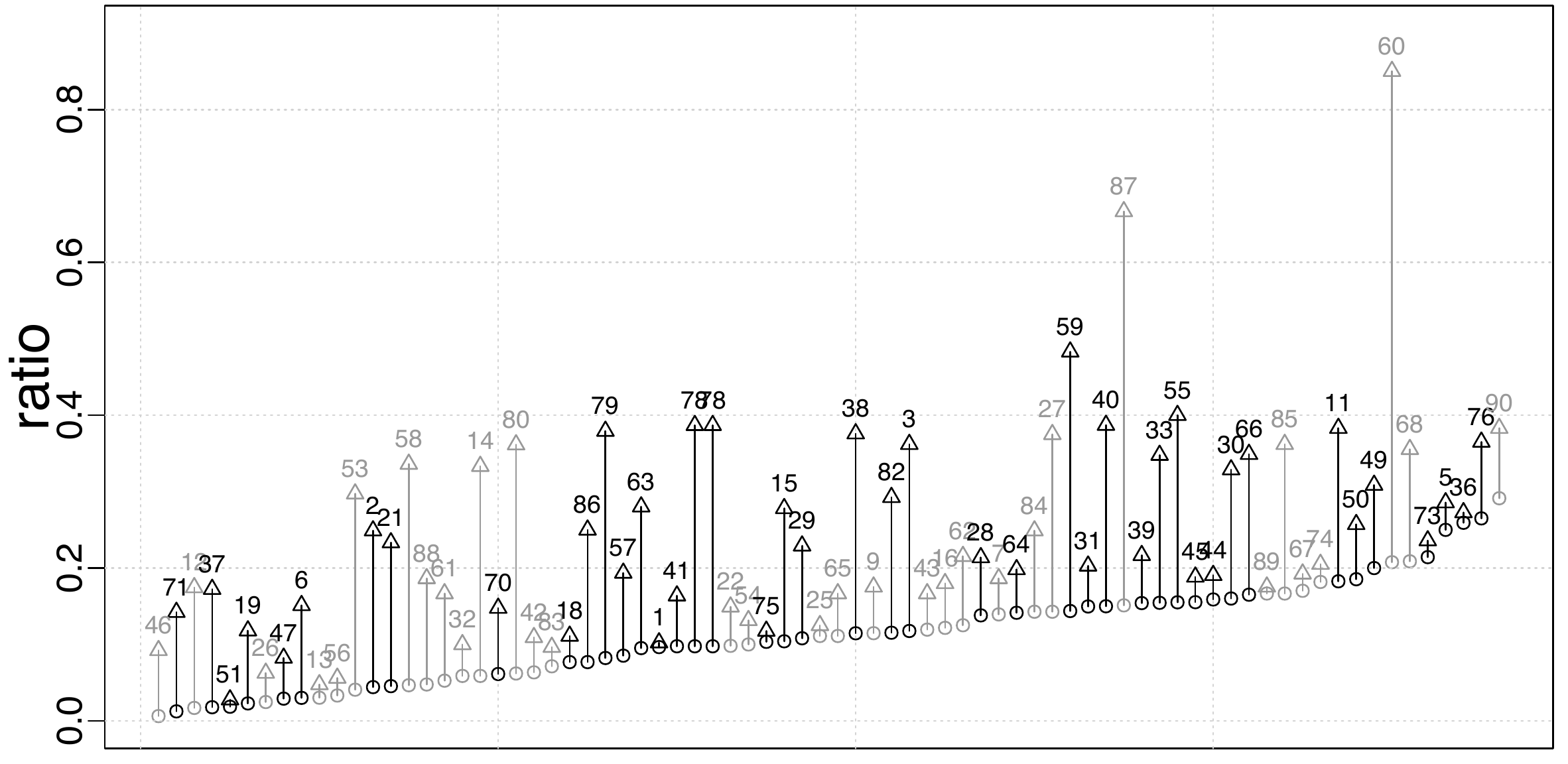}\label{fig:pairs-wp}}
\squeezeup 
\caption{Visualization of the variation of additional variables within and between sampled twins}\label{fig:pairss}
\end{figure*}

\textbf{Wordpress use}. 
Previous studies suggest that domains with popular content management systems (CMS) in general have higher odds of being compromised for phishing attacks \cite{vasek2015hacking}.
Therefore, we hypothesize that the more Wordpress websites a provider host, the higher is the chance of being a target of phishing attacks.
We use this as a proxy of the extent to which Wordpress as a popular and targeted software is used per provider. 
To collect data on Wordpress installations per provider, we randomly sample 2\,\% of the domains of each provider in $\mathcal{R}$.
This results in 1,398,928 domains.
We then use WPScan, a black box Wordpress vulnerability scanner developed and supported by Sucuri, to determine if a domain uses Wordpress \cite{wpscan}.
The variable Wordpress use is calculated by taking the ratio of scanned domains that have Wordpress installation over all scanned domains excluding those that we were unable to scan. 

To raise confidence in the selection of twins and facilitate the interpretation of estimated coefficients, in Figure \ref{fig:pairss} we show a collection of plots visualizing the variance within and between twins for the additional variables.
Circles display the value of the provider with lower value in a twin, and are the basis for sorting all twins horizontally. 
Triangles display the respective value of the provider with higher value. 

In general, the plots suggest a broad enough coverage of data points in the twins.
More specifically, as Figure~\ref{fig:pairs-ict} demonstrates, variance of the ICT development index decreases with increasing the base level (The provider shown with circle, those with lower value in a twin).
This is probably due the concentration of hosting providers (and hence twins) in a few large and highly developed countries, as witnessed in the country pairings displayed in Figure~\ref{fig:pairs_cc}.
The popularity index exhibits moderate differences at all levels (Fig.~\ref{fig:pairs-alexa}). 
This indicates that the sampling strategy accounts well for the heterogeneity in the size of hosting providers, which is also included in the index calculation. 
At the same time, the remaining variation allows for the statistical identification of potentially influencing factors within the twins. 
Also time in business is very ``healthy'' in this regard, with a smaller difference for values between providers in twins which are in business since the \texttt{.com} ages (Figure~\ref{fig:pairs-tib}).
The differences in price are rather small and exhibit occasional spikes (Figure~\ref{fig:pairs-price}). 
This may reflect the generally low cost of the cheapest package of one of the twins, differences in business models at the spikes, and potential issues related to comparing US dollar amounts among countries with very different labor and infrastructure costs. Finally, the Wordpress use also shows a good mix of variation within and between twins, increasing the chance of extracting a meaningful signal if the indicator has explanatory power (Figure~\ref{fig:pairs-wp}).

The general coverage of the rich data points $\mathcal{R}$ of the population $\mathcal{T}$ is best assessed by comparing the descriptive statistics of the four explanatory variables and the response variables, which are available for the entire population. This information is included in the lower half of Table~\ref{tab:sum_vars}.

\section{Modeling Phishing Counts}
\label{sec:regsize}

We now propose a statistical model to analyze to what extent the structural properties of hosting providers can explain the concentration of  abuse in their networks, for the case of phishing.

\subsection{Regression model}
\label{subsec:model_specification}

Abuse is measured by counting abuse events per provider. 
These counts consists of non-negative integers that arise from a direct observation of a point process.
In a minimal model, the underlying process is assumed to be stationary and homogeneous, with i.i.d.\ arrival times for abuse events and thus can be modeled with a Poisson distribution.

Let us define our response variable $Y_i$ as the number of \emph{abused} second-level domains hosted by provider $i$, for $i=1,\ldots,n$, with $n$ being the total number of hosting providers. Let  
$Y_i$ follow a Poisson distribution with parameter $\lambda\geq 0$,
\squeezeup 
\begin{equation}
Pr\left\{Y_i=k \right\}= \frac{{e^{ - \lambda_i } \lambda_i^k }}{{k!}}, \; \forall k \in  \mathbb{N} \cup \{0\}.
\end{equation}
The Poisson distribution has equal mean and variance $E\left[Y_i \right] = var\left[Y_i\right] = \lambda_i$. 


In the log-linear version of the general linear model (GLM), $\lambda_i$ is modeled as:
\squeezeup 
\begin{equation}
\ln(\lambda_i)= \beta_0 + \bm{x}'_i\bm{\beta} =  \beta_0 + \sum^k_{j=1} x_{ij} \beta_j ,
\label{eq:poisson_model}
\squeezeup 
\end{equation}
where  $\beta_0$ is the intercept, $x_{ij}$, $j=1,\ldots,k$, are explanatory variables representing the structural properties that drive the response variable $Y_i$, and $\beta_j$ are parameters to be estimated with maximum likelihood (ML). Statistical hypothesis tests can tell if a parameter $\beta_j$ significantly differs from zero. If the  null hypothesis is rejected, the corresponding explanatory variable is considered influential and the parameter's sign and magnitude can be interpreted.

\subsection{Model Goodness of fit}
\label{subsec:model_fit}

The fitted values produced by inserting the ML estimates $\hat{\bm{\beta}}$ into Eq.~\eqref{eq:poisson_model} will not match the values of the phishing data perfectly, chiefly because the data are realizations and the fitted values are parameters of Poisson distributions. 
The discrepancy between the model and the data is a measure of the inadequacy of the model. Several measures exist to assess the goodness of fit of GLMs such as Log-likelihood, Akaike Information Criterion (AIC), dispersion parameter of the Poisson model and R-squared. Here we discuss a few of them that are  more specific to our way of using the Poisson model.

\begin{table*}[!htbp] \centering 
\squeezeup 
\tbl{GLM for count of phishing domains in APWG for all the hosting providers\label{tab:regm_sizes} }{%
\resizebox{\linewidth}{!}{
\begin{tabular}{@{\extracolsep{5pt}}lrrrrr} 
\\[-1.8ex]\hline 
\hline \\[-1.8ex] 
 & \multicolumn{5}{c}{Response Variable: Count of phishing domains} \\ 
\cline{2-6} 
\\[-1.8ex] & \multicolumn{5}{c}{Poisson with Log Link Function} \\ 
\\[-1.8ex] & (1) & (2) & (3) & (4) & (5)\\ 
\hline \\[-1.8ex] 
 Number of assigned IP addresses &  & 1.186$^{***}$ & $-$1.665$^{***}$ & $-$0.728$^{***}$ & $-$0.768$^{***}$ \\ 
 &  & (0.002) & (0.006) & (0.006) & (0.006) \\ 
  & & & & & \\ 
 Number of IP addresses hosting domains&  &  & 3.623$^{***}$ & 1.104$^{***}$ & 1.570$^{***}$ \\ 
  &  &  & (0.006) & (0.008) & (0.010) \\ 
  & & & & & \\ 
 Number of hosted domains &  &  &  & 1.686$^{***}$ & 1.238$^{***}$ \\ 
  &  &  &  & (0.004) & (0.006) \\ 
  & & & & & \\ 
 Percentage of domains hosted   &  &  &  &  & 0.027$^{***}$ \\ 
on shared IP addresses  &  &  &  &  & (0.0003) \\ 
  & & & & & \\ 
 Constant & $-$0.122$^{***}$ & $-$3.594$^{***}$ & $-$2.732$^{***}$ & $-$5.072$^{***}$ & $-$6.755$^{***}$ \\ 
  & (0.005) & (0.010) & (0.011) & (0.014) & (0.024) \\ 
  & & & & & \\ 
\hline \\[-1.8ex] 
Observations & 45,358 & 45,358 & 45,358 & 45,358 & 45,358 \\ 
Log Likelihood & $-$223,113.400 & $-$514,546.600 & $-$236,442.400 & $-$117,601.700 & $-$111,570.800 \\ 
Akaike Inf. Crit. & 446,228.800 & 1,029,097.000 & 472,890.800 & 235,211.400 & 223,151.700 \\ 
\cline{2-6}
\\[-1.8ex] 
Dispersion &2934.775 & 619.708 & 554.695 & 12.149 &13.166 \\
Pseudo $R^2$ &  & 0.221 & 0.648 & 0.831 & 0.841 \\ 
\hline 
\hline \\[-1.8ex]
\textit{Note:}  & \multicolumn{5}{r}{$^{*}$p$<$0.05; $^{**}$p$<$0.01; $^{***}$p$<$0.001} \\ 
 & \multicolumn{5}{r}{Standard errors in brackets} \\ 
 \end{tabular} }}
 \squeezeup 
 \end{table*} 

\textbf{Over-dispersion}
\label{subsec:dispersion}
Recall that the Poisson model assumes equal mean and variance for
the response variable, that is $var\left[Y_i\right] = \phi E\left[Y_i\right] =\phi \lambda_i $, with $\phi\stackrel{!}{=}1$, where $\phi$ is a dispersion parameter.
However, this assumption is often ``violated'' in practice; that is, a likelihood function which leaves $\phi$ as a parameter to be estimated ($\hat\phi$) fits the data much better.
In case of heterogeneous count variables, $\hat\phi>1$ indicates signs of over-dispersion, which can be interpreted as unobserved heterogeneity in terms of a missing structural factor that leads to concentrations of observable events.

One might approach a model's over-dispersion by starting from a Poisson model and adding a multiplicative random effect to represent unobserved heterogeneity. This leads to a negative binomial GLM.
However, even if both parameters of the assumed negative binomial distribution are correctly specified, if the distribution of the response variable is not in fact the negative binomial, then the maximum-likelihood estimator becomes inconsistent \cite{cameron1990regression}. 

The literature suggests that in the absence of a precise mechanism that produces the over-dispersion, it is safe to assume $var(Y_i) =\phi \lambda_i$, for positive values of $\phi$.
This approach is generally considered robust since even significant deviations have only a minor effect on the fitted values, their standard errors, confidence intervals, and $p$-values of hypothesis tests \cite{heinzl2003pseudo}.
Moreover, over-dispersion, as a sign of unobserved heterogeneity, is exactly what we are interested in, in this paper: answering the question to what extent explanatory factors can explain differences between abuse levels of hosting providers. 
Over-dispersion it is an indicator for structural variance in our model. Any attempt to compensate it with the choice of more complex distribution functions, such as negative binomial or zero-inflated Poisson, may hide the very signal we are looking for.

The dispersion parameter $\phi$ of a Poisson regression model is calculated using the chi-square statistics $\chi^{2}$ divided by degrees of freedom, as it is more robust against outliers \cite{mittlbock2002calculating}. 
\vspace*{-5pt}
\begin{equation}
\hat{\phi} = \frac{\chi^{2}}{(n-k-1)} = \sum_{i} \frac{{\left(y_i-\hat{\lambda_{i}}\right)}^2}{\hat{\lambda_{i}} \cdot (n-k-1)},
\label{eq:disp}
\end{equation}
with $n$ being the number of observations and $k$ the number of coefficients.
$ \bm{y} = (y_{1},...,y_{n})'$ are the observed values of the response variable;  $\hat{\lambda} = (\hat{\lambda}_{1},..., \hat{\lambda}_{n})'$ are the corresponding predicted values under the fitted model, respectively.

\textbf{Pseudo R-Squared}
\label{subsec:pseudo-r}
A popular measure to assess the fraction of variance explained by a linear model is the $R$-squared statistic. Similar statistics that take values between 0 (not better than intercept-only model) and 1 (perfect fit) have been proposed for GLMs and are called \emph{pseudo} $R$-Squared.
According to \cite{heinzl2003pseudo}, a pseudo $R$-Squared measure for Poisson models that takes the effect of over-dispersion into account is given by
\begin{equation}
R^{2} = 1- \frac{D(\bm{y},\hat{\bm{\lambda}}) + k \cdot \hat{\phi}}{D(\bm{y},\bar{Y})},
\squeezeup 
\end{equation}

where $D(\bm{y},\hat{\bm{\lambda}})$ is the deviance of the fitted model, 
$D(\bm{y},\bar{Y})$ is the deviance of the intercept-only model, $\hat{\phi}$ is the estimated dispersion (Eq.~\eqref{eq:disp}), $k$ is the number of covariates fitted, (excluding intercept) and $\bar{Y} = \frac{1}{n} \sum_{i=1}^{n} y_{i}$. 

\subsection{Model specifications}
\label{subsec:mod_spec}
\noindent After selecting the proper regression model and discussing goodness of fit measures,
we choose to fit different specifications of the model with a step-wise inclusion of the variables that capture providers' structural properties. For each of the variables, we hypothesize the direction of its relation with phishing counts.

We expect that the number of phishing counts increases as the `Number of hosted domains' and `Number of IP addresses hosting domains' increase. Both variables are correlated and measure some aspects of the size of a provider. We may expect that the coefficient sizes decline if both enter the model together, but it is unlikely that one of them becomes redundant.
In case of `Number of assigned IP addresses', the assumption is slightly different since the more assigned IP addresses does not necessarily mean that the provider uses them for web hosting.
In contrast, we found that the business model of providers with larger assigned IP space is closer to a broadband provider who uses only a very small portion of its assigned space for web-hosting. 
Accordingly, since phishing attack -- as an instance of abuse -- directly depends on web-hosting, we expect providers with large `Number of assigned IP addresses', to have less phishing events in a specification where size is already controlled for with the two other variables.

In addition, note that our log-transformation of the top three variables shown in Table \ref{tab:regm_sizes}, perfectly matches with the log-link function of the Poisson model.

We expect that the variable `Percentage of domains hosted on shared IP addresses' correlates positively with the phishing counts of providers due to commonly known vulnerabilities of shared hosting services \cite{nikiforakis2011abusing,apwg2014}, assuming that attackers would go for targets that are easier to compromise.

\subsection{Estimation results}
\label{subsec:model_results}

We construct several models by performing a step-wise inclusion of the explanatory variables explained in Section \ref{subsec:mod_spec}. 
A summary of these regression models is presented in Table \ref{tab:regm_sizes}.
The table contains 5 models with estimated coefficients of explanatory variables in each model. Each coefficient demonstrates the amount of variance in phishing counts that a variable can explain given other variables in the model.

As we move from model~(2) to (5), we aim to improve our models in terms of goodness of fit metrics explained in Section \ref{subsec:model_fit}.
Likewise, the pseudo $R$-squared values increase when explanatory variables are added from model~(2) to (5). 
More specifically, with the four size and business-related variables, we are able to explain 84\,\% of the variance in abuse counts across 45,358 hosting providers with simple structural properties of providers.
One should note that the dispersion parameter $\hat{\phi}$ across the models has been reduced from 2934.77 for the intercept-only model to 13.17 in model~(5).
In addition to other factors, the significance and signs of the estimated coefficient for explanatory variables do not change between model~(3) and model~(5), which further indicates we can safely take model~(5) as our final model. 

The results in model~(5) clearly suggests that the number of phishing sites increases as the `Number of hosted domains' and `Number of IP addresses hosting domains' increases. One should note that these two variables together capture the \textit{attack surface} of the hosting providers for the case of phishing attacks, the best. Hence, the results demonstrates that for larger attack surfaces, there are more phishing websites.

As hypothesized, `Percentage of domains hosted on shared IP addresses' show a statistically significant relation with the abuse counts, indicating that having more domains on shared servers make a provider more exposed to phishing attacks.
The `Number of assigned IP addresses' shows a statistically significant negative relation with the abuse count, as expected when controlling for size.
As pointed out earlier, a manual inspection of the providers with large `Number of assigned IP addresses' suggests that they are mostly broadband providers who offer web hosting only as a very small part of their business. Therefore, the negative sign of `Percentage of domains hosted on shared IP addresses' works in line with our hypothesis of having more web hosting domains and IPs as attack surface, increases the number of phishing victims.

In addition, the coefficients and pseudo $R$-squared values in the models~(2) to (5) further confirm the point we made earlier in the introduction of the paper that a simple bi-variate correlation or a \textit{naive} normalization of abuse with one size metric, while neglecting other size properties, cannot comprehensively explain the variance in abuse counts. 

Taking model~(2) as an example, the value of estimated coefficient for `Number of assigned IP addresses' suggest that 
increasing this variable by one unit (i.e., one decimal order of magnitude), multiplies the number of expected abuse counts by $e^{1.186}$ = 3.273.
However, a \textit{naive} normalization of abuse (abuse counts divided by network size) would have assumed a coefficient equal to 1 for the `Number of assigned IP addresses'. 
In the multivariate models, several size indicators offset each other, making the estimation more precise.
In addition, in model~(2) where `Number of assigned IP addresses' is the only size variable, we are only able to explain 22\% of the variance in phishing counts whereas by adding three other size metrics, the explained variance is improved up to 84\%.

\subsection{Quantitative interpretation}
To illustrate the economic significance of the parameters in the fitted model~(5), we familiarize the reader with three scenarios. 
The scenarios are based on hosting provider groups discussed in~ \cite{nomshosting2016}, which is medium, small and large hosting providers.

\begin{figure}[h!]
\squeezeup
 \centering
\includegraphics[width=0.5\linewidth]{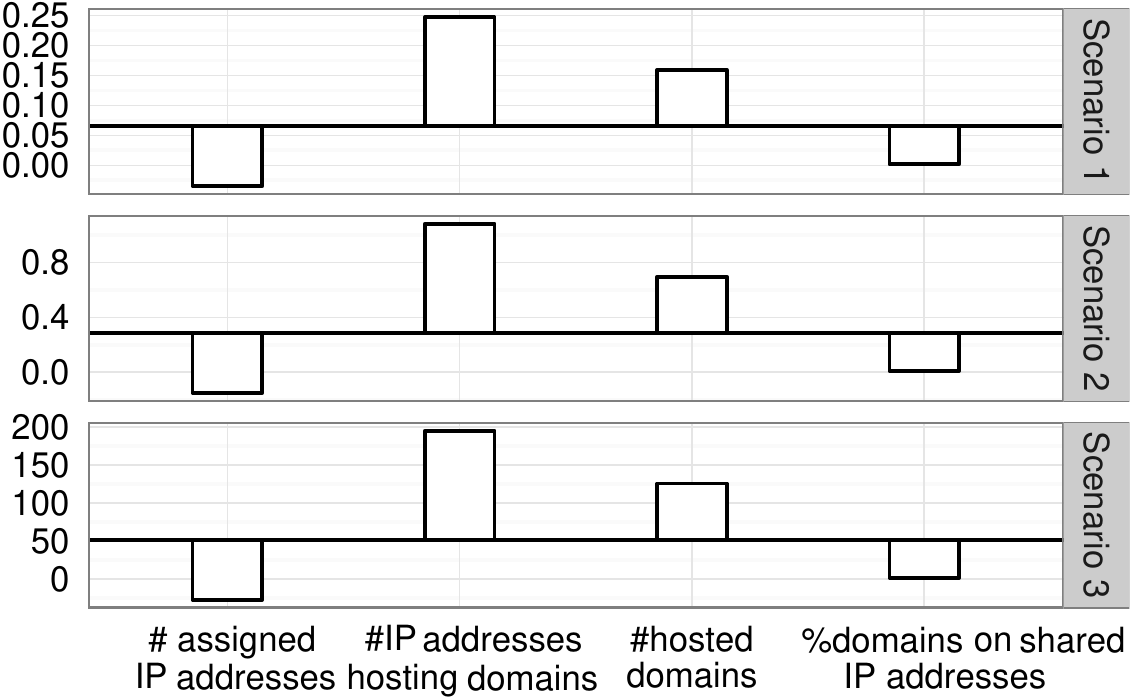}
\caption{Partial effects of one unit increase of predictors on expected phishing counts in model~(5) of Table \ref{tab:regm_sizes} (mind the different scales)}
\label{fig:scenarios}
\squeezeup
\end{figure}

\squeezeup For each scenario, we show the partial effect of changes in the abuse counts given each of the explanatory variables (see Figure~\ref{fig:scenarios}).

In the first scenario (scenario 1), all explanatory variables are set at their median as a baseline situation (see Figure~\ref{fig:scenarios}).
In scenario 2, the baseline shows a typical small shared hosting provider with a small number of assigned IP addresses (0.47: $\approx 10^{0.47}$ IPs assigned), a small number of IP addresses used for hosting domains (0.47: $\approx 10^{0.47}$ IPs ), a small number of hosted domains (1.95: $\approx 10^{1.95}$ domains) and a high percentage of domains hosted on shared IP addresses (100\%).

In scenario 3, the baseline situation indicates a large web hosting provider with huge number assigned IP address space (6.85: $\approx 10^{6.85}$, large IP address space used for hosting domains (5.67: $\approx 10^{5.67}$ IPs assigned), large amount of hosted domain names (5.68: $\approx 10^{5.68}$ domains) and very small percentage of domains hosted on shared IP addresses (0.48\%).
Apparently, this is common case for web hosting providers that are mostly offering dedicated services \cite{nomshosting2016}.

The bars for each of the scenarios in Figure~\ref{fig:scenarios} illustrate the change in the expected count of abuse events for a unit change in each of the explanatory variables, while holding the others constant.
Given the coefficients for the explanatory variables in Model~(5) in Table~\ref{tab:regm_sizes}, one can easily observe that changes in `Number of IP addresses hosting domains' while other variables are constant, changes the phishing counts the most, while the effect of one unit change in `Percentage of domains hosted on shared IP addresses' on the phishing counts is small although statistically significant.

\begin{figure}[h!]
 \centering
\squeezeup 
\includegraphics[width=0.5\linewidth]{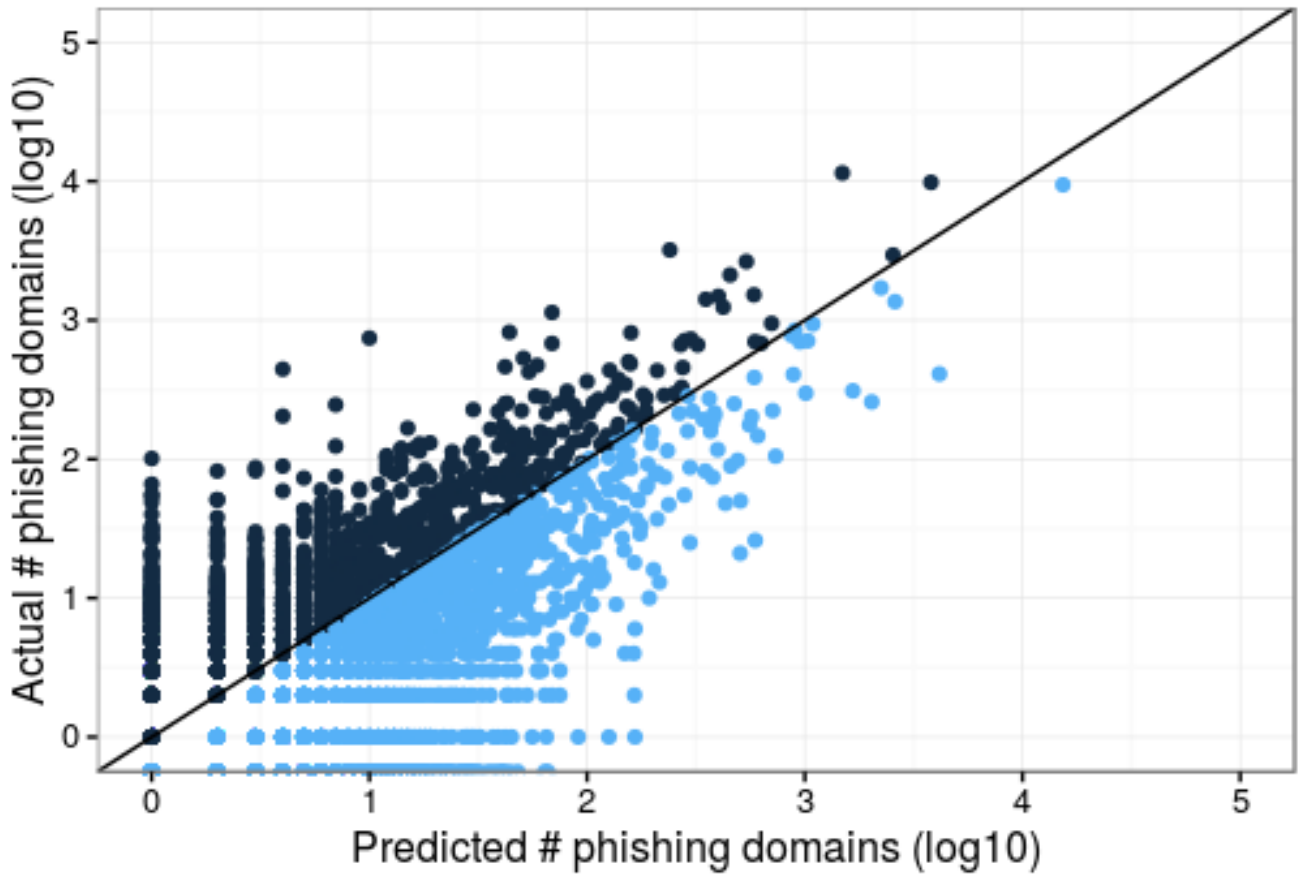}
\label{fig:obs_vs_pred}
\caption{Observed and predicted number of phishing domains for Model~(5) in Table \ref{tab:regm_sizes}}
\end{figure}

Obviously, evaluating how effective providers are in keeping phishing sites off their networks is important for developing policies and best practices. Absolute concentrations of abuse are still useful, of course. Even if they reflect structural factors, those providers could be asked to go beyond the security practices of the industry, because of their prominent position. Such a call is less plausible, however, if attackers can easily switch among the thousands of providers, as has been found in other work \cite{levchenko2011click}.

Rather than looking at absolute counts of abuse alone, measuring the amount of abuse relative to the providers' structural properties adds valuable information. Figure \ref{fig:obs_vs_pred} visualizes the difference between actual phishing counts and the counts predicted by our model. Providers below the line are performing better than average and the distance tells us by how much.

\section{Additional provider structural properties}
\label{sec:pairs}

In model~(5) of Table~\ref{tab:regm_sizes}, 
we see that around 84\% of the variance in phishing counts is explained by some of the structural properties of providers, namely, four variables related to size and business model of the hosting providers. 

In this section we continue with interpreting regression models, this time on a sample of the rich data points (set $\mathcal{R}$) as described in Section~\ref{subsec:sampling}. Recall that we need to adjust the estimation method by introducing two sets of fixed-effects, (i) for level differences between statistical twins and (ii) for level differences between countries. Fixed effects take away known linear dependence in the residuals. This is essential to obtain accurate test statistics (which assume independent residuals). Sources of dependence are within twins due to the selection strategy and within countries due to the inclusion of country-level variables.

We define the model with (one set of) fixed-effects as:
\begin{equation}
\label{eq:mod_eq_pairs}
\ln({\lambda}_{ij}) = \beta x_{ij} + \dots + \delta_i,
\squeezeup 
\end{equation}

where $\beta$ is the estimated coefficient for $x_{ij}$, $x_{ij}$ are explanatory variables collected for hosting providers in set $\mathcal{R}$ and $\delta_{i}$ is the ``fixed-effect'' parameter \cite{cameron2013regression}.
Subscript $i$ refers to different twins and $j \in \{1,2\}$ refers to different measurements within each twin, i.e., the same variable measured at different hosting providers belonging to the same twin.

The model uses variables explained in Section~\ref{subsec:data_sample} as predictors.
We add two fixed-effects to the model -- twin and country -- by fitting a separate dummy variable as a predictor for each class of each variable. 
The twin fixed effect controls for the bias introduced by selecting similar samples. 
The country fixed effect prevents undue dependence in the residuals if country-level predictors are included.
In addition, in case of missing values per explanatory variable, we perform a list-wise exclusion on a twin, if one of providers in is missing. This further reduces the number of pairs per model depending on missing values of the included variables.

With fixed effects, the definition of pseudo $R$-squared requires some reflection. It is possible to use an intercept-only baseline, which results in artificially high pseudo $R$-squared values because the level differences of the fixed effects are counted as ``explained''. A more conservative measure is to use the fixed-effects-only model as baseline.
Table \ref{tab:reg_pairs} shows the summary of our results.

\begin{table}[!htbp] \centering 
\tbl{GLM for count phishing domains in APWG for the ``statistical twins''  \label{tab:reg_pairs}  }{%
     \resizebox{0.6\linewidth}{!}{
\begin{tabular}{@{\extracolsep{5pt}}lrrrrr} 
\\[-1.8ex]\hline 
\hline \\[-2.1ex] 
 & \multicolumn{5}{c}{\thead{Response Variable \\ Count of phishing domains}} \\ 
\cline{2-6} 
\\[-1.8ex] & \multicolumn{5}{c}{Poisson with Log Link Function} \\ 
\\[-1.8ex] & (1) & (2) & (3) & (4) & (5)\\ 
\hline \\[-1.8ex] 
  Price per year &  &  &  & 0.0003 & $-$0.007$^{***}$ \\ 
  &  &  &  & (0.0002) & (0.001) \\ 
  & & & & & \\
 Popularity index &  &  & 0.001$^{***}$ & 0.02$^{***}$ & 0.1$^{***}$ \\ 
(in thousands) &  &  & (0.000) & (0.002) & (0.01) \\ 
& & & & & \\ 
  Time in business  &  &  & $-$0.017$^{*}$ & $-$0.059$^{***}$ & 0.015 \\ 
  &  &  & (0.007) & (0.005) & (0.012) \\ 
  & & & & & \\ 
ICT dev. index &  &  & 0.951$^{***}$ &  & $-$165.065 \\ 
  &  &  & (0.214) &  & ($>10^{3}$) \\ 
  & & & & & \\ 
Wordpress use  &  &  &  & 5.858$^{***}$ & 2.203$^{***}$ \\ 
  &  &  &  & (0.271) & (0.450) \\ 
  & & & & & \\ 
Twin fixed-effects & Yes & Yes & Yes & Yes & Yes\\ 
Country fixed-effects & No & Yes & Yes & No & Yes  \\ 
  & & & & \\ 
\hline \\[-1.8ex] 
\hline \\[-1.8ex] 
Observations & 210 & 210 & 180 & 84 & 82 \\ 
Log Likelihood & $-$2,783.157 & $-$1,111.825 & $-$966.249 & $-$795.838 & $-$249.780 \\ 
Akaike Inf. Crit. & 5,776.315 & 2,521.650 & 2,192.499 & 1,683.677 & 641.560 \\ 
 \cline{2-6}
\\[-1.8ex] 
Dispersion & 40.133 & 25.770 & 27.352 & 31.554 & 11.243\\ 
Pseudo $R^2$ &  &  &-0.055 & 0.625 & 0.772 \\ 
Total pseudo $R^2$ & 0.974 & 0.991  & 0.992 & 0.966  & 0.995 \\ 
\hline 
\hline \\[-1.8ex] 
\textit{Note:}  & \multicolumn{5}{r}{$^{*}$p$<$0.05; $^{**}$p$<$0.01; $^{***}$p$<$0.001} \\ 
 & \multicolumn{5}{r}{Standard errors in brackets} \\ 
\end{tabular} }}
\squeezeup 
\end{table} 

Table \ref{tab:reg_pairs} contains two baseline models (model~(1) and (2)). 
While the former only models twins as a fixed-effect, the later models both twins and countries as fixed effects.
Model~(3) broadens model~(2) by fitting more explanatory variables with fewer missing values. 
In model~(4) we add all the explanatory variables collected for the set of rich datapoints ($\mathcal{R}$), except for the ICT development index, having only twins as a fixed effect.
In addition to variables in model~(4), in model~(5) we add ICT development index, setting both twin and country as fixed effects.

The regression results in Table \ref{tab:reg_pairs} indicate that by including both twins and country as fixed effects, keeping in mind that the datapoints in set $\mathcal{R}$ are already homogeneous in terms of other variables, we are able to explain around 77\% of the variance in phishing counts for the twins (pseudo $R$-squared value). Note that this variance is the 77\% of the 16\% unexplained variance that is remained in model~(5) of Table \ref{tab:regm_sizes}, for the full population of providers. The results further highlights the importance of  accounting for other --non-size related-- structural properties of providers, while explaining the variance in concentration of abuse. 

Even though we have around hundred fixed-effects, we still get very clear and significant results for the main effects.
In addition, by moving from model~(1) to model~(5) in Table \ref{tab:reg_pairs}, we are reducing the amount of unexplained heterogeneity (model's dispersion) from 40.133 to 11.243.

As hypothesized, we see a significant negative relation between the price of hosting and phishing counts in model~(5).
That is, if we increase price by one unit holding the other variables constant, the phishing count will be multiplied by $e^{(-0.007)}=0.99$ which means that the cheaper a service is, the more the hosting provider is prone to phishing attacks.
Interestingly, variable `Price per year' shows a different relation in model~(4) where we do not control for cross-country differences.
This change is expected, however, as properties of hosting markets in different countries can differ a lot, which eventually influences the cost structure of the country in regards to hosting services.
In addition, the cost of a hosting plan is relative to the economy of the provider's country. Conversion of prices in a specific country to USD, if not controlled for the country differences, can be very crude.
Variables `Wordpress use' and `Popularity index' also show a significant positive relation with phishing count indicating that more Wordpress sites per provider translates to more phishing attacks, which is quite logical considering the fact that the majority of phishing sites are compromised machines.
One unit increase in `Wordpress use' while holding other variables constant, multiplies the phishing counts by $e^{(2.203)}=9.052$.
Similarly, the more popular websites of a provider is on average, the most that provider is a victim of phishing attacks.

For `ICT development index', we observe both a sign and significance change from model~(3) to model~(5).
This can be understood by looking back at the distribution of twins in Fig.~\ref{fig:pairs-alexa}, where the gray color marks the twins that were excluded from Model~(5) due to missing values.
From the figure it is visible that the 100 observations that were excluded because of missing price information are clustered among lower developed countries, thereby removing the variance needed to find the effect of ICT development.
The effect is also easily observable in Model~(3). Without the price variable,  `ICT development index' shows a significant and positive relation.

Now the question is, to what extent does our sample reflect the population? 
Looking back to our sampling strategy, in model~(5) of Table \ref{tab:regm_sizes}, we have a model that explained 84\% of heterogeneity; 
so looking for neighbors in the projection of model~(5) increases the chance of getting twins that are very comparable for all the factors that we cannot observe and are already somewhat correlated to size measures.
This means that the enriched data points contain more valuable information than others from the total population.
However, since instead of a totally random sample, we are creating twins of providers, our targeted sampling strategy might introduce possible biases.
In order to deal with that bias, we make an assumption that the bias is linear in the modeling domain, i.e., can be captured by linear fixed effects parameters.
In Section \ref{sec:measure_error}, we further perform additional cross-validations, to check for possible biases our methodology might have introduced.

\section{Robustness Checks}
\label{sec:measure_error}

During the course of our analysis, we pointed out a few assumptions that we have made, most notably about the impact of the deviations from the Poisson model due to over-dispersion; and about the impact of measurement errors in the abuse data.
In this section we address these two concerns. 
Regarding the first assumption, we use a simulation study to perform a robustness check on our size estimates. To check for possible errors in the phishing data, we cross validate with another data source.

\subsection{Size estimates}
Assume attack events \textbf{are} Poisson and the only structural factor that affects them is the size, i.e., the attack surface. 
In the absence of perfect information of the true attack surface, it can only be approximated in practice through the size variables we used in Table \ref{tab:regm_sizes}. Now we would like to study: to which extent are deviations from Poisson observable only by using the imperfect size estimations?
The precise steps towards estimating new models for phishing abuse counts using imperfect size estimations are as follows:

\begin{figure}[ht!]
\squeezeup 
\begin{minipage}{.5\linewidth}
\centering
\includegraphics[width=0.8\linewidth]{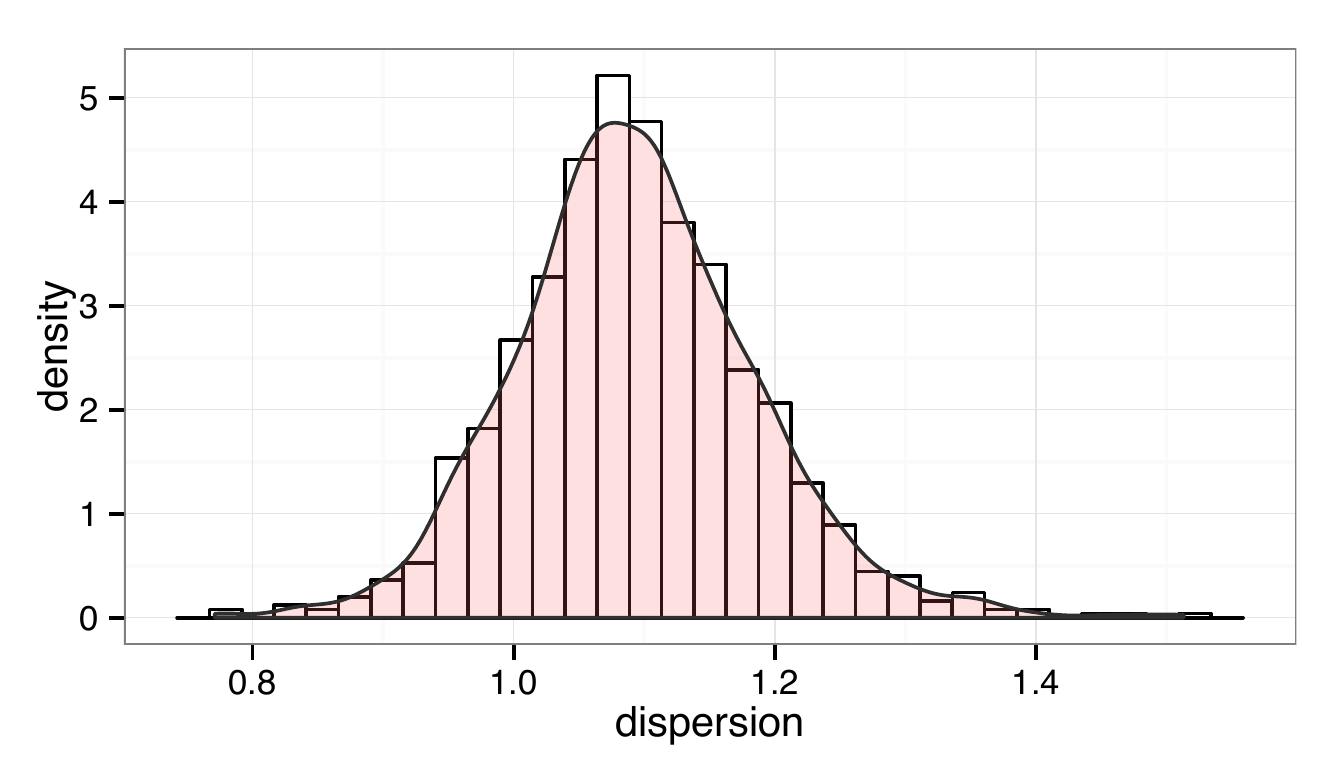}
\label{fig:disp}
\end{minipage}
\begin{minipage}{.5\linewidth}
\includegraphics[width=0.8\linewidth]{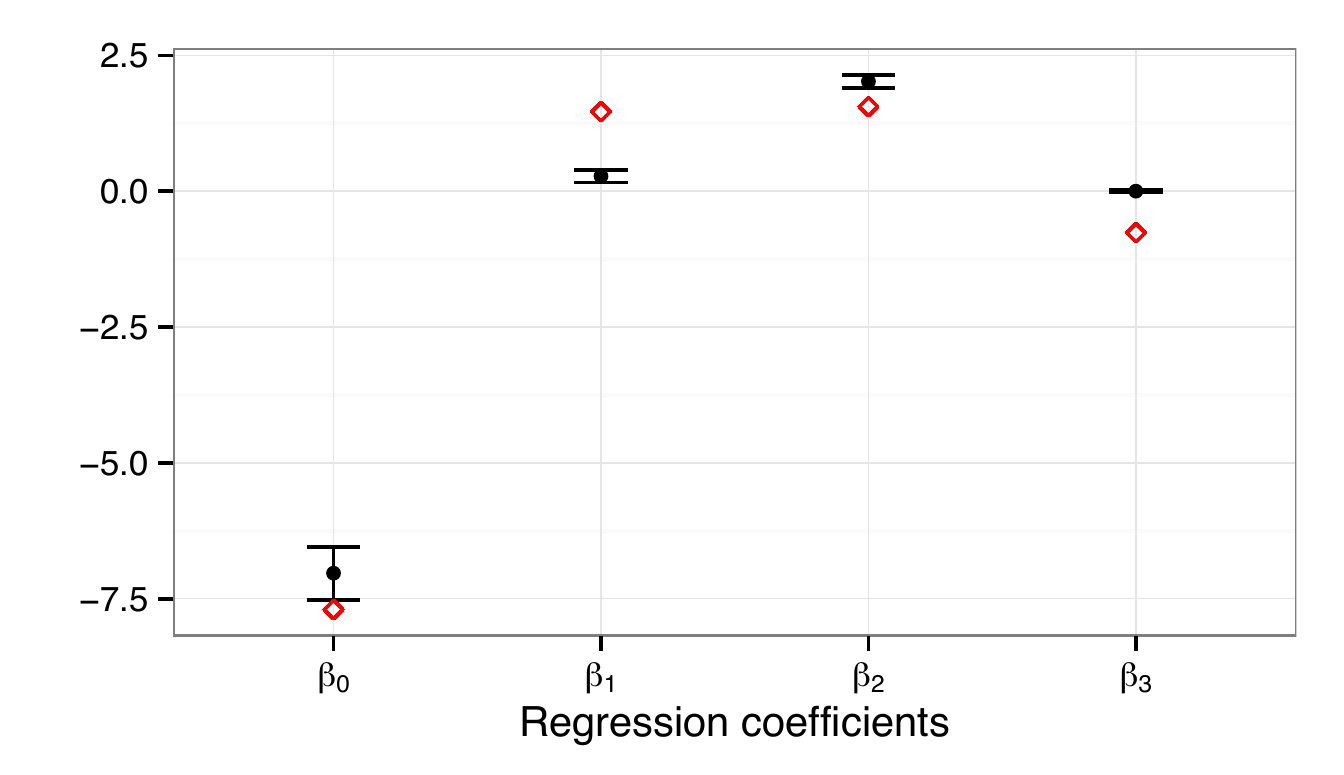}
\label{fig:coef}
\end{minipage}
\squeezeup 
\caption{(Left) Density of dispersion estimates obtained from synthetic Poisson phishing arrival with imperfect size control-(Right) Error bars of the simulated regression coefficients}
\end{figure}

We generate a \textit{true size} variable that is following normal distribution using mean and standard deviation of variable `Number of hosted domains'.
We then simulate attack events --\textit{hits}-- where the phishing counts follow a Poisson distribution ($Pois(\lambda_{sim})$).
where $\lambda_{sim}$ is the mean number of phishing domains. 
We then build the simulated size variables used in model~(3) of Table \ref{tab:regm_sizes}, namely, `Number of hosted domains', `Number of IP addresses hosting domains', `Number of the assigned IP addresses' by adding random noise ($N(\mu_{f_i},\sigma_{f_i}), \forall i\in\{1..3\}$) to our true size variable. 
$\mu_{f_i}$ and $\sigma_{f_i}$ are estimated using the mean and standard deviation of the corresponding explanatory size variables. 

We generate 1,000 times the synthetic data representing both the size and the dependent variables, and model them using the GLM regression specified in model (4) (see Table~\ref{tab:regm_sizes}). 
Then we calculate the dispersion parameter for each one of the simulated models using Eq.\eqref{eq:disp}.
Figure~\ref{fig:disp} shows the density distribution of the dispersion parameter for each of the Poisson models using simulated size measures. 
The model indicates that the dispersion parameter is on average greater than $1$. The dispersion parameter value is far from the dispersion parameter that we obtained using our real dataset in model~(4) of Table~\ref{tab:regm_sizes}.
This however is expected since the real size estimates are far from being perfect and contain several outliers.
Moreover, the over-dispersion in simulated size variables indicates that some part of the over-dispersion in model~(4) of Table~\ref{tab:regm_sizes} -- probably not everything -- can be  attributed to the approximate measurement of the size estimates. 
Finally, Figure~\ref{fig:coef} displays the coefficients of the 1,000 model fits as error bars. The red diamonds are the coefficients obtained with the full Poisson model.
The coefficients follow the same trend as in the model given the over-dispersion, which validates its robustness.

\subsection{Phishing data}
Given the known limitations using blacklists as abuse data, e.g., the lack of overlap \cite{metcalf2015blacklist}, we ran our approach against an alternative data source: the Cyscon phishing data. This allows us to see how sensitive our results are to specific data sources. 

Similar to before, we model arrival rate of phishing counts using a Poisson GLM with log-link function with size and business model predictors as used in model~(4) and model~(5) of Table \ref{tab:regm_sizes}. The result of the two final models is displayed in Table \ref{tab:cys_sizes}.
We then model the statistical twins in the set of rich datapoints  $\mathcal{R}$, having two sets of fixed effects for differences between twins and countries.
The results are shown in Table \ref{tab:cys_pairs}.
Model~(1) and (2) contain the same explanatory variables as Model~(4) and (5) of Table \ref{tab:reg_pairs}.
Interestingly, and reassuringly, the resulting estimated coefficients and significance levels for both of the analysis are quite similar to those of the model with APWG data.

\begin{table}[!htb] 
	\begin{minipage}{.48\linewidth}
      \tbl{GLM for count of phishing domains in Cyscon data for all  the providers\label{tab:cys_sizes} }{%
\centering 
        \resizebox{\linewidth}{!}{%
       \begin{tabular}{lrr} 
   		 \\[-1.8ex]\hline 
        \hline \\[-2.1ex] 
       & \multicolumn{2}{c}{\thead{Response Variable\\ Count of phishing domains}} \\ 
      \cline{2-3}
       \\[-1.8ex] 
      \\[-1.8ex] & \multicolumn{2}{c}{Poisson-Log Link Function} \\ 
      \\[-1.8ex] & (1) & (2)\\ 
      \hline \\[-1.8ex] 
       & & \\ [-1ex]
       Number of assigned IPs & $-$0.719$^{***}$ & $-$0.776$^{***}$ \\ 
        & (0.011) & (0.012) \\ 
         & & \\ [-1ex]
       Number of IPs hosting domains & 1.170$^{***}$ & 1.751$^{***}$ \\ 
        & (0.014) & (0.018) \\ 
         & & \\ [-1ex]
       Number of hosted domains & 1.663$^{***}$ & 1.115$^{***}$ \\ 
        & (0.007) & (0.011) \\ 
         & & \\ [-1ex]
       Percentage of domains hosted on  &  & 0.033$^{***}$ \\ 
      shared IPs  &  & (0.001) \\ 
       & & \\ [-1ex]
       Constant & $-$6.432$^{***}$ & $-$8.488$^{***}$ \\ 
        & (0.026) & (0.045) \\ 
        & & \\ 
      \hline \\[-1.8ex] 
       & & \\ 
      Observations & 45,358 & 45,358 \\ 
      Log Likelihood & $-$49,763.500 & $-$47,208.470 \\ 
      Akaike Inf. Crit. & 99,535.010 & 94,426.950 \\ 
      \cline{2-3}
       & & \\ [-1ex]
      Dispersion &20.153 &17.444 \\ 
      Pseudo $R^2$ & 0.791 &0.803\\
       \hline 
      \hline \\[-1.8ex] 
      \textit{Note:}  & \multicolumn{2}{r}{$^{*}$p$<$0.1; $^{**}$p$<$0.05; $^{***}$p$<$0.01} \\ 
       & \multicolumn{2}{r}{Standard errors in brackets} \\ 
      \end{tabular} 
      }}
	\end{minipage}  
	\begin{minipage}{.48\linewidth}
    \tbl{GLM for count of phishing domains in Cyscon data for the providers in the ``statistical twins''
  \label{tab:cys_pairs} }{%
      \resizebox{0.97\linewidth}{!}{%
        \begin{tabular}{lrr} 
          \squeezeup 
        \\[-1.8ex]\hline 
        \hline \\[-2.1ex] 
         & \multicolumn{2}{c}{\thead{Response Variable\\Count of p	hishing domains}} \\ 
        \cline{2-3} 
        \\[-1.8ex] & \multicolumn{2}{c}{Poisson with Log Link F				unction} \\ 
        \\[-1.8ex] & (1) & (2)\\ 
        \hline \\[-1.8ex] 
         Price per year & 0.0003 & $-$0.012$^{***}$ \\ 
          & (0.0003) & (0.002) \\ 
          & & \\ [-1ex]
         Popularity index (in thousands) & 0.02$^{***}$ & 0.1$^{***}$ \\ 
          & (0.004) & (0.01) \\ 
          & & \\ [-1ex]
         Time in business  & $-$0.048$^{***}$ & 0.004 \\ 
          & (0.010) & (0.037) \\ 
          & & \\ [-1ex]
          ICT dev. index &  & 13.610 \\ 
          &  & ($>10^3$) \\ 
          & & \\ [-1ex]
         Wordpress use & 7.848$^{***}$ & 3.079$^{**}$ \\ 
          & (0.583) & (1.125) \\ 
          & & \\ [-1ex]
         Pair Fixed-Effect &Yes& Yes \\ 
        Country Fixed-Effect & No & Yes \\ 
          & & \\ 
        \hline \\[-1.8ex] 
        Observations & 84 & 82 \\ 
        Log Likelihood & $-$476.818 & $-$145.676 \\ 
        Akaike Inf. Crit. & 1,045.635 & 433.352 \\ 
        \cline{2-3}
        \\[-1.8ex] 
        Dispersion & 20.712 & 2.993\\
        Fixed-effects Pseudo $R^2$ & 0.538& 0.889 \\ 
        Total Pseudo $R^2$ & 0.970 & 0.987 \\ 
        \hline 
        \hline \\[-1.8ex] 
        \textit{Note:}  & \multicolumn{2}{r}{$^{*}$p$<$0.05; $^{**}$p$<$0.01; $^{***}$p$<$0.001} \\ 
         & \multicolumn{2}{r}{Standard errors in brackets} \\ 
        \end{tabular} }}
  \end{minipage}
\end{table}

\section{Conclusions and Outlook}
\label{sec:conclusions}

The core question of this paper is: to what extent are abuse levels determined by the security efforts of individual providers versus being a function of structural properties of the industry? We now summarize our findings and discuss the implications of the results.

We reduced the abuse attribution error, by empirically studying the population of hosting providers for the first time, which are defined based on actors who operate the IP space rather than technical resources such as ASes.
Next, we advanced the existing work that uses simple regression analysis by taking disentangle a variety of factors and errors at work in abuse counts.
By building several GLM models for phishing abuse counts as a response variable, we demonstrated that a handful of providers' structural properties, such as domain names space size, IP space size used for web-hosting and size of their shared hosting business, can already account for 84\% of the variation in phishing counts. 
These variables are easily measurable at scale and capture the `attack surface' of providers along with aspects of their business model.

Additionally, we presented previously unstudied factors, such as pricing, time-in-business of a provider, and the amount of Wordpress sites per provider and others collected via a tailored sampling approach, can explain a further 77\% of the remaining variation in phishing.

Our findings suggest that abuse rates for phishing reflect an overall bad harvest rather than being driven by rotten apples who don't care about security. 
Our approach enhances more informative comparison of providers. In other words, it generates comparative abuse metrics by controlling for the structural differences among providers. Such relative metrics are more suited to evaluate countermeasures than absolute counts and concentrations of abuse.


When structural factors are so dominant in driving abuse, it undermines the common narrative to call for better security practices of apparently under-performing actors or for even more radical interventions, such as sanctions. However, our findings do not limit the action space for policy. Quite the contrary: data-driven policy could try to improve the factors identified as influential, e.\,g., require higher security standards at providers who host more popular websites. Furthermore, relative abuse metrics can, in themselves, incentivize better security \cite{Konte2015aswatch,hecybersecurity2015}. 
In sum: throwing out a few rotten apples might appear more tractable, but producing a better harvest is definitely possible.  We discuss below how future work can directly apply our approach to answer these questions for other forms of hosting abuse.   

Here, we should also acknowledge several limitations of our work. First, our method is geared towards identifying the main explanatory factors in the population of hosting providers. Our conclusions should not be misinterpreted as evidence that there are no misbehaving hosting providers, only that their impact on the population of phishing incidents is surprisingly limited.

Second, the presence of certain unobserved factors, including attacker behavior, is also a limitation of this work. 
That being said, we have considerably reduced the likelihood of these being major factors in the abuse patterns, as witnessed by the variance explained by the structural properties.

Third, certain structural factors might indirectly capture some information about security efforts. 
One could argue, for example, that the price model chosen by a provider might also signal how many resources are available for security. 
On the other hand, the fact that 84\% of the variance is explained by purely technical structural properties, unrelated to price, suggests that also this impact is limited. Only more direct observation of provider security efforts can establish how it is related to price. 
We have ``measured'' effort only as a latent factor in the unexplained variance of our model. Future work could try to measure it more directly, via surveys or experiments with abuse notifications and takedown speed~\cite{Li16,cetin15}. 

A final limitation is that our empirical evidence is specific to phishing. Our modeling approach is agnostic to the type abuse, however. The independent variables and model design are not specific to phishing.  
Future work can use our approach to identify the impact of these structural and business model factors on other types of abuse in the hosting market. For some sources, like drive-by-download sites, we expect similar patterns. For other, more idiosyncratic types of abuse, like long-living botnet C\&C servers, we might expect different patterns. There, we might indeed find that rotten apples drive the abuse rates, rather than a bad overall harvest.


\bibliographystyle{ACM-Reference-Format-Journals}
\bibliography{Bibliography}

\end{document}